\documentclass{aims}
\usepackage{amsmath}
  \usepackage{paralist}
  \usepackage{graphics} 
  \usepackage{epsfig} 
\usepackage{graphicx}  \usepackage{epstopdf}
 \usepackage[colorlinks=true]{hyperref}
   \usepackage[T1]{fontenc} 
\usepackage{amsmath}

\usepackage{amssymb}
\usepackage{bm}
\usepackage{graphicx}
\usepackage{xcolor}
\usepackage{caption}
\usepackage{subcaption}
\usepackage{appendix}
\usepackage{bigints}
\usepackage{yfonts}
\usepackage{bbold}

\hypersetup{urlcolor=blue, citecolor=red}

\def\currentvolume{}
 \def\currentissue{}
  \def\currentyear{}
   \def\currentmonth{}
    \def\ppages{pp.1-32}

\theoremstyle{definition}

\newcommand{\be}{\begin{equation}}
\newcommand{\ee}{\end{equation}}
\newcommand{\ba}{\begin{array}}
\newcommand{\ea}{\end{array}}
\newcommand{\bqa}{\begin{eqnarray}}
\newcommand{\eqa}{\end{eqnarray}}
\newcommand{\bea}{\begin{eqnarray}}
\newcommand{\eea}{\end{eqnarray}}

\title[NHD and QID of AB Equations] 
      {Non-holonomic and Quasi-integrable deformations of the AB Equations}

\author[Kumar Abhinav, Indranil Mukherjee and Partha Guha]{}


 \email{kumar.abh@mahidol.ac.th}
 \email{indranil.m11@gmail.com}
 \email{partha.guha@ku.ac.ae}

\thanks{$^*$ Corresponding author: Kumar Abhinav}

\begin{document}
\maketitle
\centerline{\scshape Kumar Abhinav$^*$}
\medskip
{\footnotesize
 \centerline{Centre for Theoretical Physics \& Natural Philosophy}
   \centerline{``Nakhonsawan Studiorum for Advanced Studies", Mahidol University,}
   \centerline{Nakhonsawan Campus, Phayuha Khiri, Nakhonsawan 60130, Thailand}
} 

\medskip

\centerline{\scshape Indranil Mukherjee}
\medskip
{\footnotesize
 \centerline{School of Management Sciences,}
   \centerline{Maulana Abul Kalam Azad University of Technology,}
   \centerline{Haringhata, Nadia, Pin-741249, India}
}

\medskip

\centerline{\scshape Partha Guha}
\medskip
{\footnotesize
 \centerline{Department of Mathematics,}
   \centerline{Khalifa University of Science and Technology,}
\centerline{Zone-1, Main Campus, Abu Dhabi, United Arab Emirates,}
}

\bigskip


\begin{abstract}
For the first time both non-holonomic and quasi-integrable deformations are obtained for the AB system of coupled equations. The AB system models geophysical and atmospheric fluid motion along with ultra-short pulse propagation in nonlinear optics, and serves as a generalization of the well-known sine-Gordon equation. The non-holonomic deformation retains integrability subjected to higher-order differential constraints whereas the quasi-AB system, which is partially deviated from integrability, is characterized by an infinite subset of quantities (charges) that are conserved only asymptotically given the solution possesses definite space-time parity properties. Particular localized solutions to both these deformations of the AB system are obtained, some of which are qualitatively unique to the corresponding deformation, displaying similarities with physically observed excitations.
\end{abstract}

\section{Introduction}
The AB system is a coupled integrable system of differential equations that describes various physical systems. The prominent one among such systems is the baroclinic instability in atmospheric and in geophysical flows when the mean potential energy flow is converted into the perturbing kinetic energy, modelled by a two-layer system with constant but different densities having a non-zero shear velocity \cite{Dodd,Pedlosky}. The AB dynamics is further responsible for particular modulation instabilities in fluid dynamics \cite{Gibbon,Wang}, describes ultra-short optical pulse propagation in nonlinear optics \cite{Dodd,Tan} and in the case of cold gravity current represents the mesoscale gravity current transmission on a slopping bottom \cite{Mooney1996}. 

The model itself consists of two layers of immiscible homogeneous fluids confined by two pairs of horizontal and vertical friction-less planes to an infinitely long channel that rapidly rotate about a vertical axis \cite{Pedlosky,Gibbon79}. At equilibrium the lighter fluid resides over the denser and some gravitational potential energy converts into the kinetic energy of the fluid flow, contributing in the growth of any infinitesimal wave-like disturbances already present in the fluid that marks the baroclinic instability. A test wave-packet placed on the zonal sheer flow can describe the aforementioned dynamics aptly through a perturbation expansion \cite{Pedlosky,Gibbon79} described by the following pair of equations \cite{Tan,Gibbon79,Moroz}:

\bea
&&\left(\partial_T+c_1\partial_X\right)\left(\partial_T+c_2\partial_X\right){\mathcal A}=n_1{\mathcal A}-n_2{\mathcal A}{\mathcal B},\nonumber\\
&&\left(\partial_T+c_2\partial_X\right){\mathcal B}=\left(\partial_T+c_1\partial_X\right)\vert{\mathcal A}\vert^2.
\eea
Here $X$ is the coordinate along the channel and $T$ is time. The complex function ${\mathcal A}(X,T)$ represents the slow-varying wave-packet amplitude satisfying ${\mathcal A}\to 0$ as $\vert X\vert\to\infty$ and the real function ${\mathcal B}(X,T)$ depicts the subsequent modification of the basic flow. Real parameters $n_{1,2}$ are related to the ratio of the wave amplitudes in the two layers of the fluid whereas $c_1\neq c_2$ are the group velocities of the two wave modes. A different pair of functions constructed from the pair ${\mathcal A},{\mathcal B}$ as,

\be
A_0=\sqrt{2}{\mathcal A},\quad B_0=\pm 1-\frac{n_2}{\vert n_1\vert^2}{\mathcal B},\ee
aided by the variable re-definitions, 

\be
x=-\sqrt{n_2}\frac{X-c_1T}{c_1-c_2},\quad t=\frac{\vert n_1\vert^2}{\sqrt{n_2}}\frac{X-c_2T}{c_1-c_2}
\ee
lead to the canonical form of the system \cite{Gibbon79},

\be
2B_{0,x}+\left(\vert A_0\vert^2\right)_t=0,\qquad A_{0,xt}=A_0B_0;\quad B_0\in\mathbb{R}.\label{1}
\ee
Still one of the fields ($A_0(x,t)$) is complex and the other ($B_0(x,t)$) is real. The two fields are further related through the normalization condition $\vert A_{0,t}\vert^2+B_0^2=1$ following suitable scaling subjected to the expected boundary condition $B_0\to\pm1$ for $\vert x\vert\to\infty$ \cite{Gibbon79}. On substituting for $B_0$ the AB equations combine to,

\be
A_0^2\left(\vert A_0\vert^2\right)_t+2A_0A_{0,xxt}-2A_{0,x}A_{0,xt}=0,\label{5}
\ee
yielding a single higher order nonlinear differential equation.

Standard tools such as the method of Lax pair \cite{Kamchatnov}, inverse scattering transform \cite{Gibbon79}, Hirota bilinearization \cite{Wu} etc. have been employed to study the AB system of equations to establish its integrability through infinitely many conserved quantities \cite{Guo,Zhang}. The Lax pair for the AB system is given as \cite{Kamchatnov}:

\be
L=-i\lambda\sigma_3+\frac{A_0}{2}\sigma_+-\frac{A_0^*}{2}\sigma_-,\quad M=\frac{1}{4i\lambda}\left(-B_0\sigma_3+A_{0,t}\sigma_++A_{0,t}^*\sigma_-\right),\label{2}
\ee
with spectral parameter $\lambda$ and $SU(2)$ matrices:
\be
\sigma_3=\begin{pmatrix}
1 & 0\\
0 & -1
\end{pmatrix},\quad \sigma_+=\begin{pmatrix}
0 & 1\\
0 & 0
\end{pmatrix},\quad \sigma_-=\begin{pmatrix}
0 & 0\\
1 & 0
\end{pmatrix},
\ee
which leads to the AB equations through the well-known zero curvature condition:

\be
F_{tx}=L_t-M_x+\left[L,M\right]=0,
\ee
thereby ensuring integrability. Subsequently, various particular solutions of the system has been obtained. Classes of localized AB solitons \cite{Yu,Wang2,Tan} along with breather excitations \cite{Guo} and periodic solutions \cite{Tan} have been obtained. Rogue wave solutions to the AB system with higher order iteration \cite{Wang434} and their connection to baseband modulation instability \cite{Wang} were also derived. Multi-component \cite{Wu,Geng} coupled generalizations of the AB systems possess further properties such as multi-dark solitons \cite{Xie}, bright-dark solitons with mutual elastic interactions \cite{Xu} and additional nonlinear wave modes with bright-dark rogue waves \cite{Wu} which are highly indicative of the nonlinear effects \cite{Su}. Recently, converted wave solutions and their complexes in the multi-component AB system has been realized \cite{Zhang2021}.

In addition to representing the baroclinic instability in geophysical flows, the fundamental AB system of Eq.s \ref{1} further represents self-induced transparency when $A_0$ is complex \cite{Dodd,Kamchatnov} and when $A_0$ is real the system can be converted into the celebrated sine-Gordon system \cite{Dodd,Gibbon,Gibbon79}. The latter scenario is implemented through the choice $A_0=\phi_x,~B_0=\pm\cos\phi$ with $\phi\in\mathbb{R}$. The first of Eq.s \ref{1} then reduces to the sine-Gordon equation:

\be
\phi_{xt}=\pm\sin\phi,
\ee
with the second equation and the normalization relation being consistent. The sine-Gordon equation is completely integrable and play important and diverse roles both in nonlinear dynamics \cite{Dodd} and in quantum field theory \cite{Rajaraman} that include one-dimensional crystal dislocation \cite{SG01}, magnetic flux propagation across Josephson junctions \cite{SG02}, charge density waves \cite{SG04,SG05}, phonon mode excitations \cite{SG06} and deformation dynamics in DNA double helix \cite{SG07,SG08,Ivancevic} etc. Exact single and multi-soliton, breather and travelling wave solutions of the sine-Gordon equations are well-known \cite{SG09,SG10} along with some more distinct solutions coming from Darboux and B\"aklund transformations \cite{SG11,SG12} and inverse scattering \cite{SG13} of the known ones. Such well-spread utility and importance of the sine-Gordon system further emphasizes the usefulness of the AB system in analyzing physical processes.

A very crucial aspect of the integrable systems are deformations that extend their domains of study either through their generalizations or by obtaining newer systems as a result. Particularly, deformations of integrable models with physical application are of significant interest. A few deformed models of physical interest include scalar and gauge field extensions to 1+1 dimensional gravity \cite{ID01}, gauge-deformed nonlinear $\sigma$-models \cite{ID02,ID03} which are important in string theory, deformed nonlinear Schr\"odinger equation \cite{ID04,AK3,A3,A5,A6} which is widely applicable as effective theory of many physical systems, and deformation of the KdV equation \cite{ID05,Kuppershmidt,AK1,AK2,Guha1,tB,RKdV} that describes shallow hydrodynamic waves. Deformations of sine-Gordon equation \cite{A2,RSG,Barone,ID06,ID07} are also known to describe various realistic situations. As discussed above, in case of the AB system itself, though coupled versions \cite{Wu,Xie} and multi-component generalizations \cite{Xu,Su,Geng} have been studied, genuine deformations are yet to be investigated as per the best of our knowledge.

The present work examines the effects of two different types of deformations of physical importance on the AB system. The first is the Non-holonomic deformation (NHD) that preserves the integrability of the system through additional higher-order constraints, whereas the second being the Quasi-integrable deformation (QID) that deviates the system from integrability and is characterized by anomalous conservation laws. For different integrable systems both NHD \cite{AK3,Kuppershmidt,AK1,AK2,Guha1,KK,Guha2} and QID \cite{A3,A5,A6,A2,A1,A7,A4} have been obtained and studied at length. We elaborate on both these deformations in the next couple of sections. Though entirely different in nature, comparative studies of NHD and QID for physically very relevant nonlinear Schr\"odinger equation \cite{AGM2} and its generalizations \cite{AGM1} have been carried out recently. Since the AB system finds various physical implications and also serves as a generalization to the sine-Gordon equation which also is physically relevant, the corresponding NHD and QID should be of considerable interest which has not been attempted yet. 

The NHD of the AB system is found to support localized soliton-type excitations akin to the undeformed system \cite{Gibbon79}. However the known two-soliton, kink-kink and kink-anti-kink sectors of the undeformed system \cite{Gibbon79} do not find counterparts upon NHD. We further obtain an anti-kink type solution unique to the deformed sector. Such distinct solution sectors highlight the possibility of the NHD of AB system representing physical systems quite distinct to the AB system itself. Moreover a redundancy in defining the field $B_0$ through a linear shift by a local function arises under NHD at particular spectral order that suggests a possible class of solutions given a set $\left(A_0,\,B_0\right)$. The QID of the AB system is found to effect a local deviation in the single soliton sector at the leading order of expansion whereas in the two-soliton sector the effect of QID seems to be at the boundary. In both the cases the quasi-deformed one- and two-soliton structures deviate only marginally from the their undeformed counterparts \cite{Gibbon79}, suggesting strongly that the quasi-AB systems can correspond to baroclinic unstable systems having irregularities both locally and at the boundaries. Further, an infinite set of quasi-conserved quantities are obtained which regain conservation only at the spatio-temporal boundary subjected to particular space-time parity of the solution sector.

The rest of the paper is organized as follows. In Section 2 the non-holonomic deformation of the AB system has been obtained and a few particular localized solutions of the same are discussed. Only a few solutions of the undeformed AB system found to survive the NHD whereas other unique structures are obtained exclusive to the deformed system. The quasi-integrable deformation of the AB system is discussed in Section 3 obtaining quasi-deformed one- and two-soliton solutions that represent partial conservation. A $\mathbb{Z}_2$ automorphism analysis is performed establishing generic quasi-integrability of the AB system subjected to definite space-time parity properties of the solutions. We conclude in Section 4 after some discussions regarding physical relevance and utility of our results encompassing known experiments on baroclinic flows, followed by possible avenues of future investigations.

\section{NHD of the AB system}
The NHD of integrable systems is one in which the system is perturbed in such a manner that under suitable differential constraints on the perturbing function, the system retains its integrability. Karasu-Kalkani {\it et al} \cite{KK} showed that the integrable 6th order KdV equation represents a NHD of the celebrated KdV equation. The terminology `nonholonomic deformation' was used by Kupershmidt \cite{Kuppershmidt}. The N-soliton solution using inverse scattering transform and a two-fold integrable hierarchy were obtained for the NHD of the KdV equation \cite{AK1}. This work was extended to include both KdV and mKdV equations along with their symmetries, hierarchies and integrability \cite{AK2}. The NHDs of the derivative NLS and the Lenells-Fokas equations were discussed in Ref. \cite{AK3}. Therein, corresponding deformed integrable versions were obtained with some arbitrary functions of time as coefficients and subsequent solutions were found to give rise to the phenomenon of accelerating solitons with suitable choice of the time dependent coefficients. The NHD of generalized KdV type equations was taken up in Ref. \cite{Guha1} where emphasis was put on the geometrical aspect of the problem. Kupershmidt's infinite-dimensional construction was further extended to obtain the NHD of a wide class of coupled KdV systems, all of which are generated from the Euler-Poincare-Suslov flows \cite{Guha2}. The NHD of the non-local generalization of the nonlinear Schr\"odinger equation has also been studied recently \cite{MG}. 

To construct the NHD of an integrable system one starts with the corresponding Lax pair, keeping its spatial component $L$ unchanged but modifying the temporal one $M$. As a result the corresponding scattering profile remains unchanged, but the time evolution of the spectral data changes due to the deforming perturbation. To retain integrability the non-holonomic constraints have to be affine in velocities prohibiting explicit velocity dependence of the deformed dynamical equation. Such a requirement demands the spatial deformation to be exclusive to the temporal Lax component leading to velocity independence of the dynamical equation since the flatness condition does not include time derivative of the temporal component \cite{NHD1}. As a result the system can retain its integrability even after being subjected to perturbation. 

Since the AB system has diverse physical applications it is viable to seek its NHD since many systems have been known to get generalized or get related to other integrable systems through NHD \cite{AK3,AK1,AK2,Guha1,AGM1}. As a particular example, certain NHD of the NLS system were found to represent both the inhomogeneous Heisenberg spin chain and dragged vortex filament \cite{A8}. For the AB system too such a deformation could at least enhance the collective understanding of such nonlinear systems with particular solution sectors being identified. It may also be possible to identify the non-holonomically deformed AB system with some new physical system. In the following we provide a general recipe for deforming the AB system non-holonomically. It is seen that deforming the temporal component of the corresponding Lax pair with terms of zero or higher order in spectral parameter $\lambda$ do not change the AB system, {\it i. e.}, no deforming term gets added to the equations at its particular spectral order. We therefore start with a deforming extension to the temporal Lax component at ${\mathcal O}\left(\lambda^{-1}\right)$ as,
\be
M_{\text{def}}= \frac{1}{4i\lambda}(u_1\sigma_3 + v_1\sigma_{+} + w_1\sigma_{-}),\label{NN05}
\ee
where the local functions $u_1$, $v_1$ and $w_1$ will be determined in terms of $A_0$, $B_0$ and their derivatives. Using the deformed temporal Lax component in the zero-curvature equation, we obtain the non-holonomically deformed AB equations as:

\bea
&&2B_{d,x} + (\vert A_d\vert^2)_t - 2u_{1,x} = 0,\nonumber\\
&&A_{d,xt}= A_dB_d - A_du_1,\quad A^*_{d,xt}= A^*_dB_d- A^*_du_1,
\eea
with $v_1$ and $w_1$ vanishing identically. The suffix $d$ signifies the deformed solutions. In comparison to Eq.s \ref{1} an extra term each get introduced on account of the NHD which, however, is effectively equivalent to a local shift $B_0\to B_d-u_1$ and therefore the combined Eq. \ref{5} remains unaltered. Further, there is no differential constraint on $u_1$ separately. Therefore the present choice of $M_{\text{def}}$ reveals potentially an infinite number of solutions through the choice of $u_1$ that are non-holonomically equivalent, a property which can effectively simplify the solution mechanism as we will see. Since the equations are unchanged under the variable redefinition $A_0\to A_d,~B_0\to B_d-u_1$ the NHD at ${\mathcal O}\left(\lambda^{-1}\right)$ essentially maps among different sets of solutions of the undeformed and deformed AB system. It could be useful in finding solutions to the deformed AB system given a known pair $\left(A_0,\,B_0\right)$. In order to obtain non-trivial results the NHD must be extended to the order $\lambda^{-2}$ terms as,

\be
M_{\text{def}}= \frac{1}{4i\lambda}(u_1\sigma_3 + v_1\sigma_{+} + w_1\sigma_{-}) + \frac{1}{4i\lambda^2}(u_2\sigma_3 + v_2\sigma_{+} + w_2\sigma_{-}).\label{NN10}
\ee
From the zero-curvature condition, considering $\lambda$-free terms, one gets $v_1=0=w_1$ as before. At order $\lambda^{-1}$ the dynamical equations are obtained as,
\bea
&&2B_{d,x} + \left(\vert A_d\vert^2\right)_t = 2u_{1,x},\nonumber\\
&&A_{d,xt} - A_dB_d + 2iv_2 + A_du_1 = 0,\quad A^*_{d,xt} - A_d^{*}B_d - 2iw_2 + A_d^{*}u_1 = 0.\label{NN06}
\eea
The last two equations are now genuinely deformed even after the local shift $B_0\to B_d-u_1$ is considered. As the flatness condition is imposed these equations represent a {\it new} integrable system with {\it source}. At order $\lambda^{-2}$ the deforming functions get related through the constraint conditions:
\be
2u_{2,x} = A_dw_2 + A_d^{*}v_2,\quad v_{2,x} + A_du_2 = 0,\quad w_{2,x} + A_d^{*}u_2 = 0.\label{NN07}
\ee
Apparently, the normalization $\vert A_{0,t}\vert^2+B_0^2=1$ is no longer valid and is replaced by the derived condition,

\be
\vert A_{d,t}\vert^2+\left(B_d-u_1\right)^2=2i\int_x\left(A_{d,t}w_2-A_{d,t}^*v_2\right).\label{01N}
\ee
Substituting from the deformed dynamical equations the constraint conditions relate higher order differentials of the deformed solutions $A_d$ and $B_d$ as,

\bea
&&iA_d\left(A_{d,xt}-A_dB_d+A_du_1\right)_{xx}-2A_d,x\left(A_{d,xt}-A_dB_0+A_du_1\right)_x\nonumber\\
&&\qquad\qquad\quad=A_d^*\left(A_{d,xt}-A_dB_d+A_du_1\right)+A_d\left(A_{d,xt}-A_dB_d+A_du_1\right)^*,
\eea
wherein the unconstrained variable $u_1$ can be absorbed through the redefinition $B_d\to B_d-u_1$. This makes further sense as now the combination $B_d-u_1$ is real definite instead of $B_d$ alone. Since the unconstrained function $u_1$ cannot be determined it is meaningful to consider $B_d-u_1$ instead of $B_d$ as a solution to the deformed system. Being of higher differential order the above constraint restricts the solution space without hampering the dynamics. As $B_d-u_1$ is real, by taking complex conjugation of the second of Eq.s \ref{NN06} one finds $v_2^*=w_2$, which implies $u_2$ to be real too. Then from the second of Eq.s \ref{NN06} by substituting $B_d-u_1$ from the first,
 \be
A_{d,xt}+\frac{1}{2}A_d\partial_x^{-1}\left(\vert A_d\vert^2\right)_t+2iv_2= 0 .\label{NN08} 
\ee
Further, substituting for $u_2$ from the second of Eq.s \ref{NN07} in the first:
\be
2\left(\frac{w_{2,x}}{A_d^{*}}\right)_x + A_dw_2 + A_d^{*}w_2^{*} = 0.\label{NN09}
\ee 
Eq.s \ref{NN08} and \ref{NN09} can be solved simultaneously in principle to obtain deformed solutions of the AB system. As a result the deformation introduced in Eq. \ref{NN10} becomes genuine. The higher order constraint comes from Eq.s \ref{NN07} that preserves the integrability through the zero curvature condition.

\subsection{Specific Examples}
The general solutions $\left(A_d,\,B_d-u_1\right)$ of the nonholonomically deformed system for a given set of deformation functions $\left(u_1,\,u_2,\,v_2=w^*_2\right)$ can be difficult to obtain as the Eq.s \ref{NN06} and \ref{NN07} represent an inhomogeneous coupled nonlinear system. However, since $v_2$ (and thus, $w_2$) can be determined in terms of $A_d$ from Eq. \ref{NN08}, it is far convenient to look for deformed configurations subjected to a particular choice of $A_d$. Then the validity of the same can be tested against the existence of the counterpart solution $B_d-u_1$. We specifically consider the known soliton solutions of the undeformed AB system as candidate solutions to the deformed AB system. Being collective excitations, solitons are localized weakly interacting modes even in systems consisting of strongly interacting particles. As a result solitons signify high degree of symmetry with subsequent conservation laws. Therefore such solutions, if exist, represent physically robust situations which can conveniently be detected. 

\begin{figure}
\centering
\begin{subfigure}[b]{0.22\textwidth}
\centering
\includegraphics[width=\textwidth]{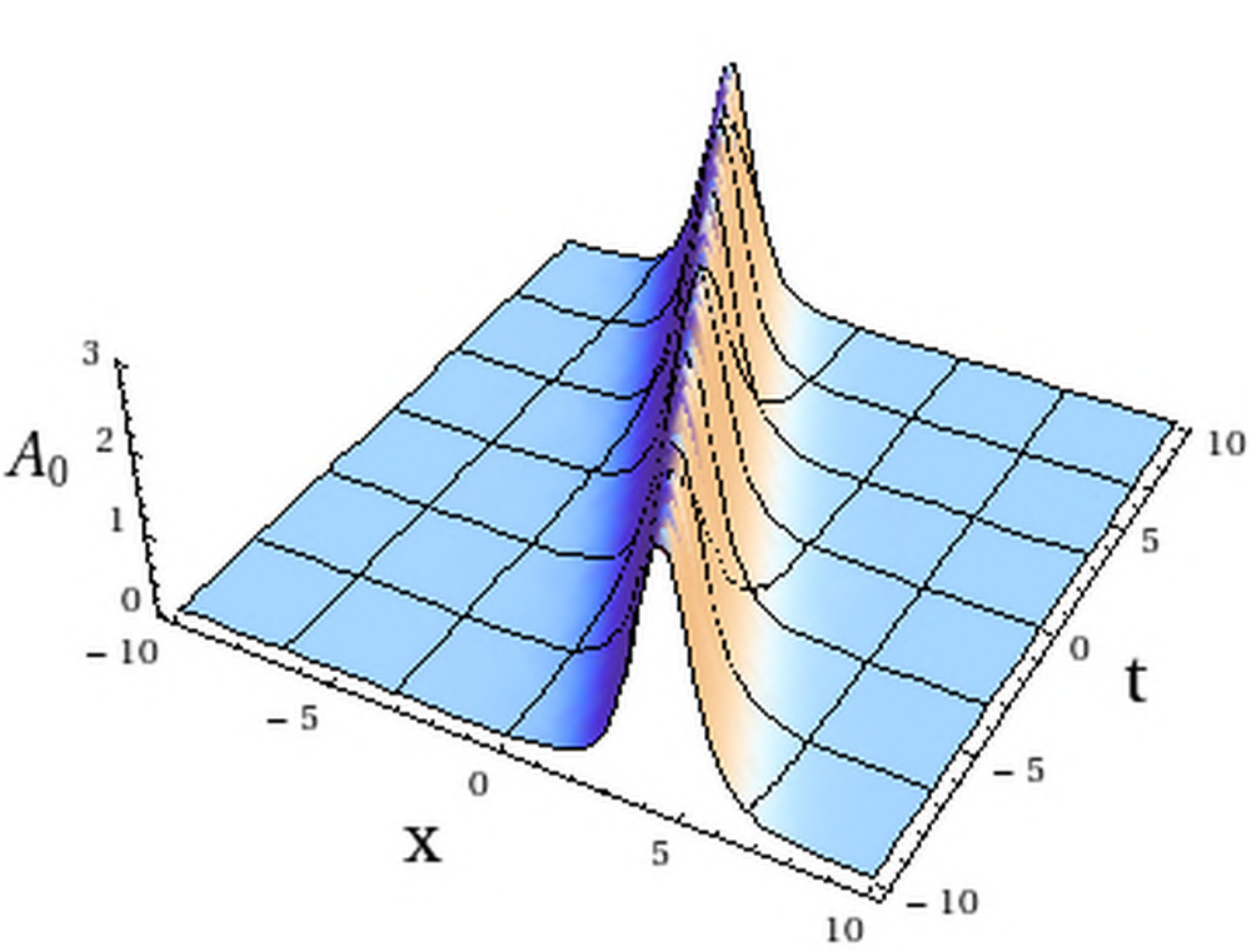}
\caption{}
\end{subfigure}
\begin{subfigure}[b]{0.22\textwidth}
\centering
\includegraphics[width=\textwidth]{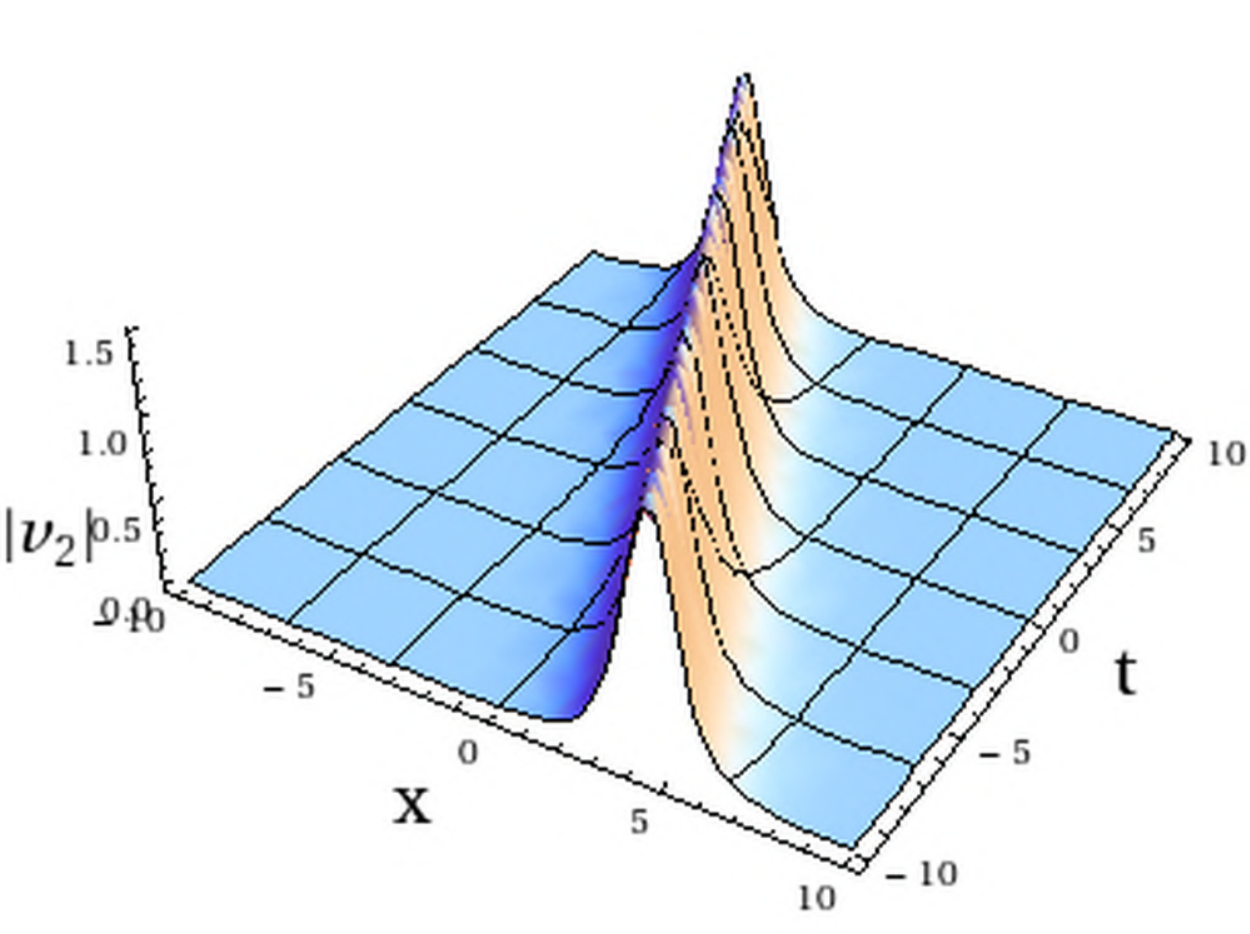}
\caption{}
\end{subfigure}
\centering
\begin{subfigure}[b]{0.22\textwidth}
\centering
\includegraphics[width=\textwidth]{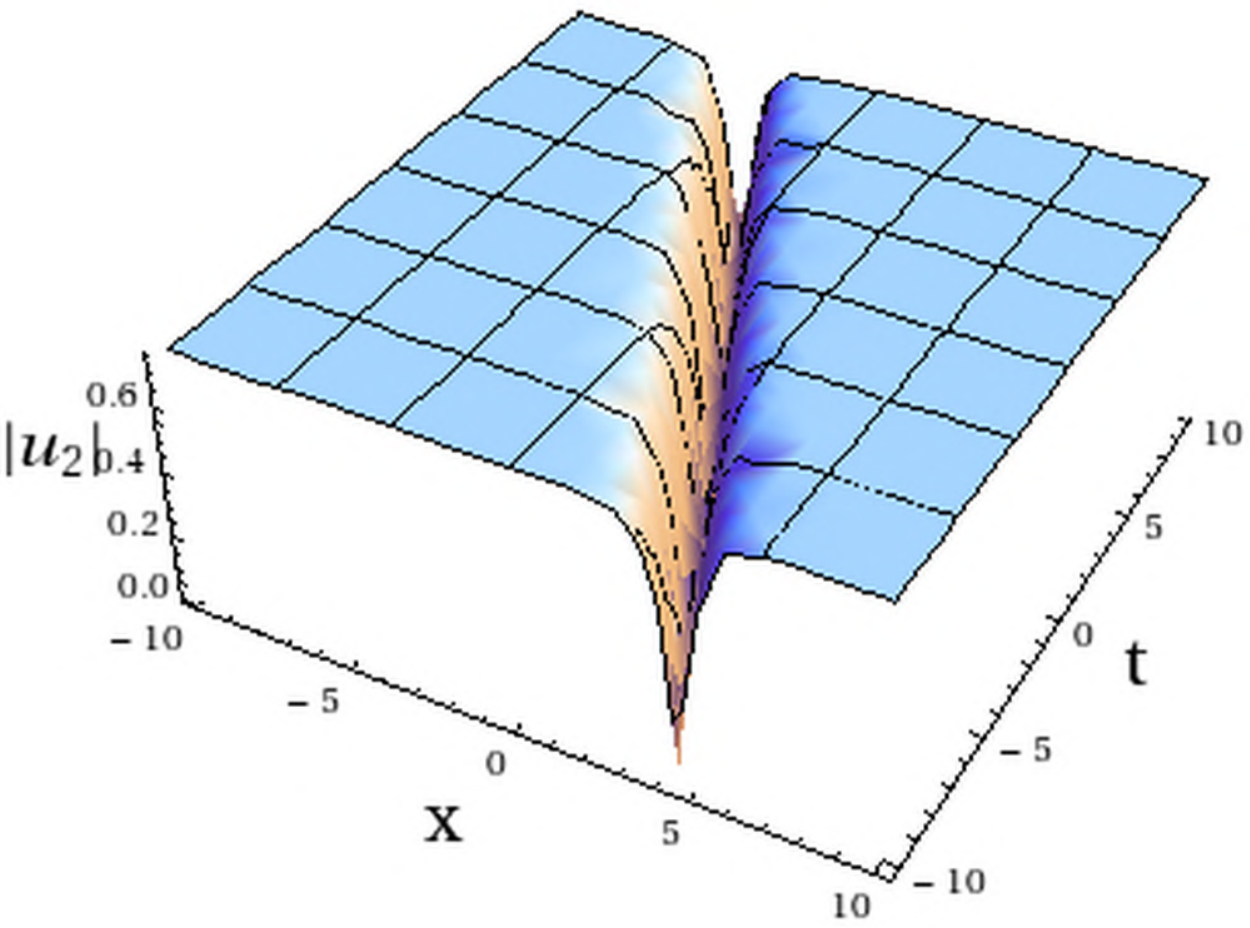}
\caption{}
\end{subfigure}
\begin{subfigure}[b]{0.22\textwidth}
\centering
\includegraphics[width=\textwidth]{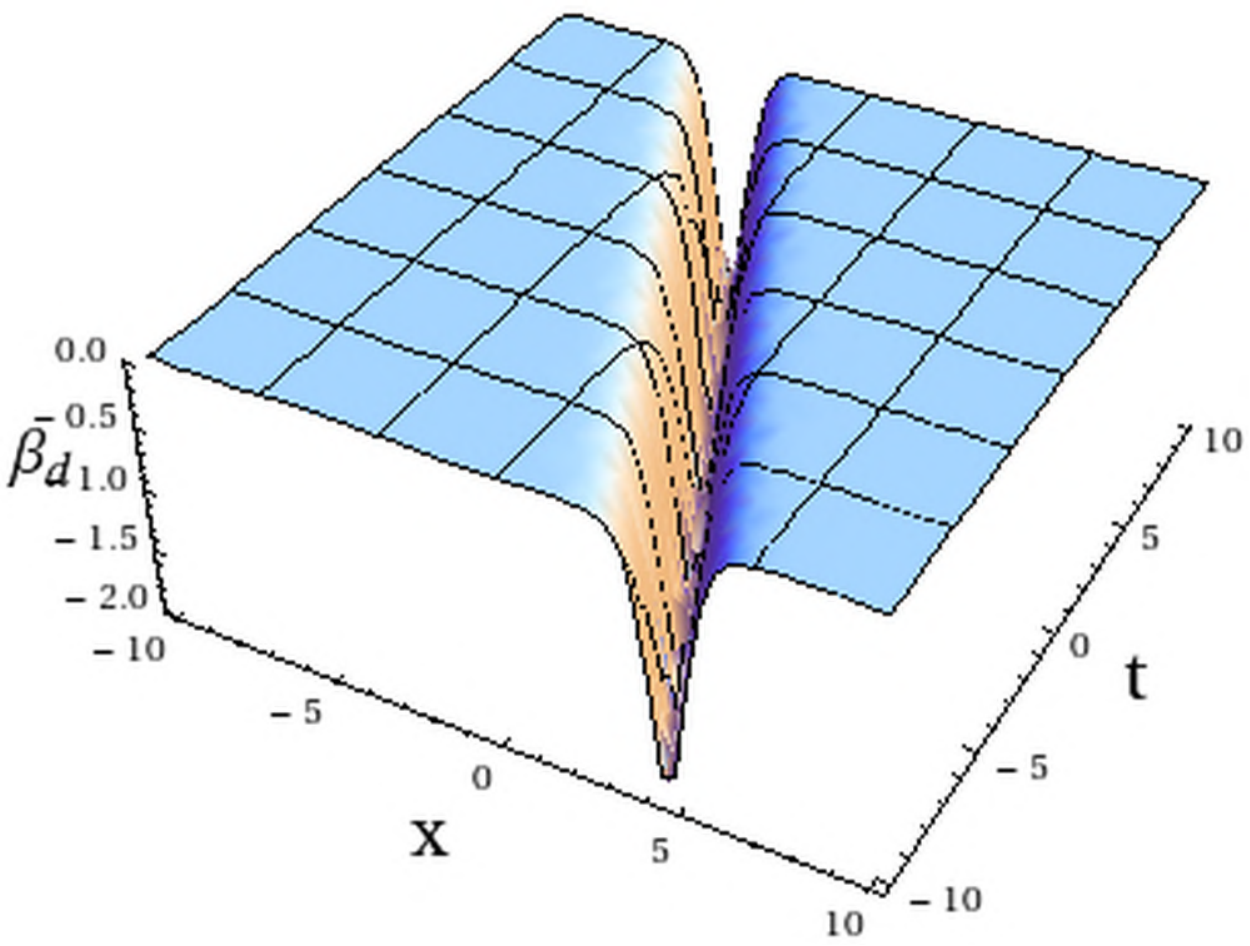}
\caption{}
\end{subfigure}
\caption{a) The deformed amplitude $A_d$ is considered to be same as the undeformed 1-soliton solution ($i\gamma=1.5$). The moduli of the corresponding complex deformation functions $v_2$ and $u_2$, which are well-localized, are shown in b) and c) respectively. d) The shifted deformed amplitude $\beta_d=B_d-u_1$ is both well-behaved and localized, validating the ansatz $A_d=A_0$ in this case.}
\label{F1Sol}
\end{figure}

To begin with, the single-soliton solution to the undeformed AB system \cite{Gibbon79} is considered,

\be
A_0=2i\gamma{\rm sech}\theta,\quad B_0=1-2{\rm sech}^2\theta;\quad\theta=i\gamma x-\frac{i}{\gamma}t+\delta;\quad\gamma\in{\mathcal I},\quad\delta\in\mathbb{R},\label{NH01}
\ee
which respects the normalization $\vert A_{0,t}\vert^2+B_0^2=1$. Let us consider the particular possibility of the deformed solution being $A_d=A_0$. Then Eq. \ref{NN08} yields,
\be
v_2=-\gamma{\rm sech}\theta=w_2^*\equiv-w_2,
\ee
leading to $u_2=-(\gamma/2)\tanh\theta$. Both the functions are well-behaved and localized. Further, 

\be
B_d-u_1=-2{\rm sech}^2\theta\equiv B_0-1,
\ee
has a valid dark soliton structure. Thus we obtain a 1-soliton solution to the nonholonomically deformed AB-system subjected to the particular set of inhomogeneities ($v_2,\,u_2$). These functions are plotted in Fig. \ref{F1Sol} using Mathematica8. In this case, one is free to choose $u_1=1$, allowing for $B_d=B_0$. Essentially the present system is a marginal deformation of the AB system and thus is expected to model some close variation to the baroclinic instability. 

\begin{figure}
\centering
\begin{subfigure}[b]{0.22\textwidth}
\centering
\includegraphics[width=\textwidth]{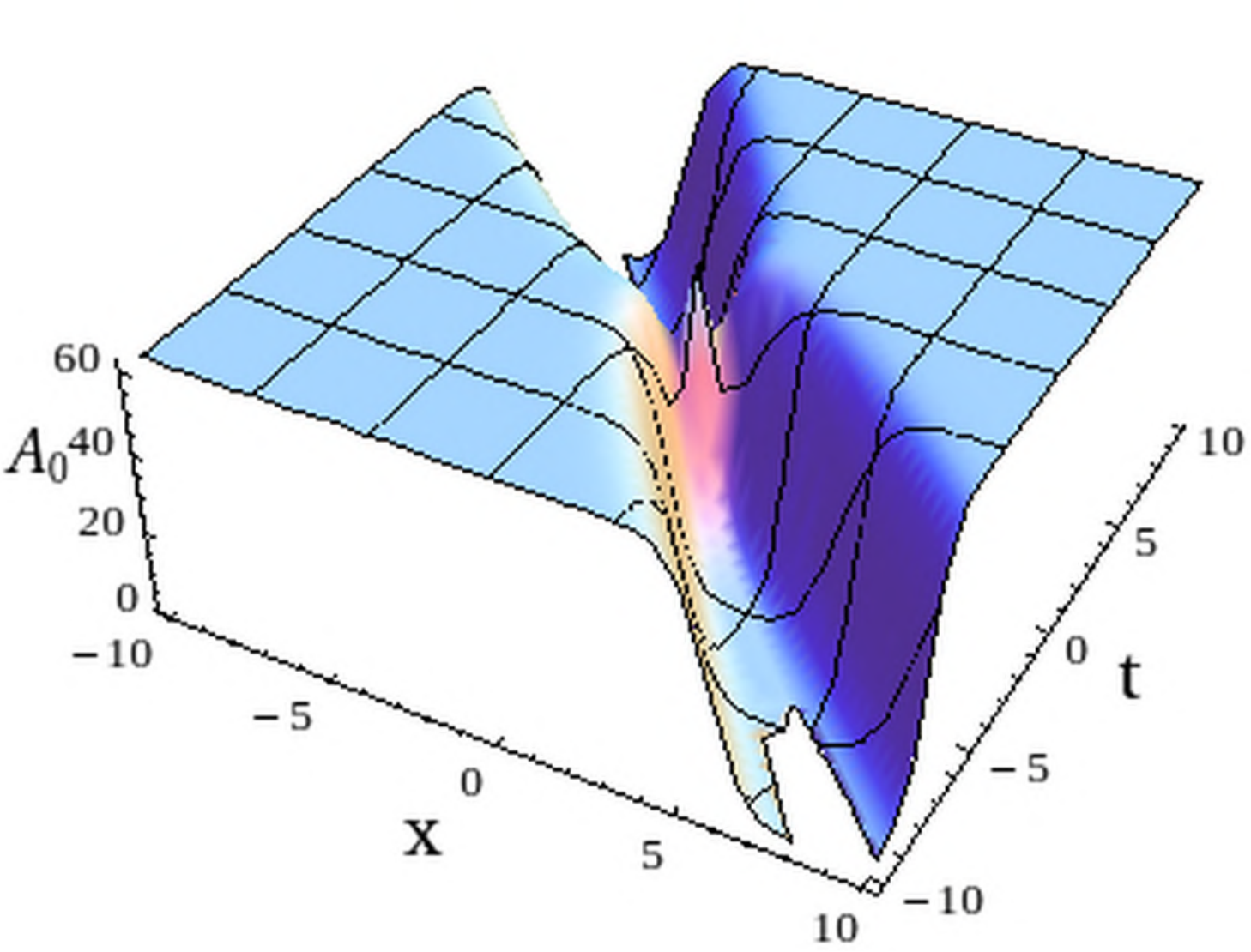}
\caption{}
\label{F1a}
\end{subfigure}
\begin{subfigure}[b]{0.22\textwidth}
\centering
\includegraphics[width=\textwidth]{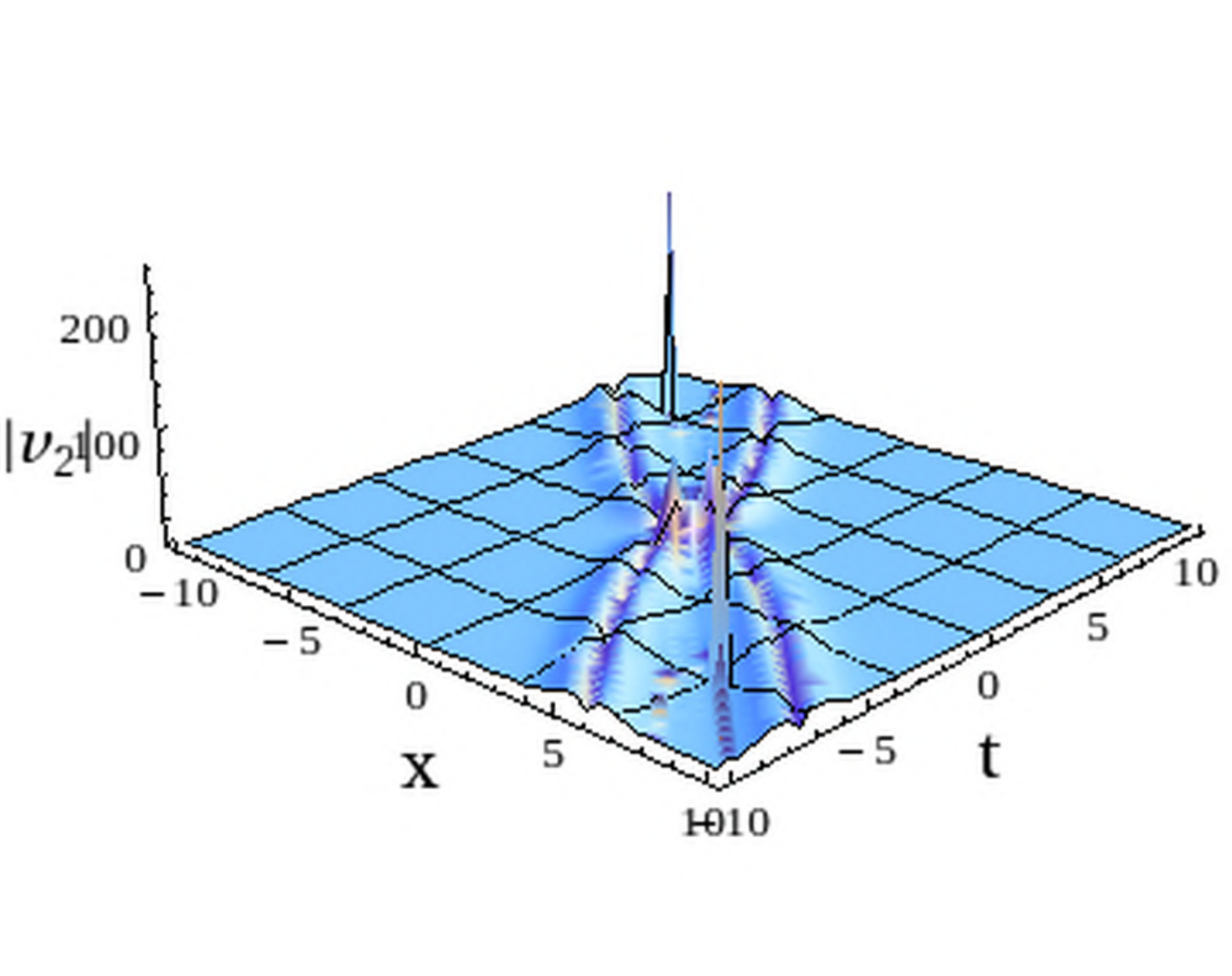}
\caption{}
\label{F1b}
\end{subfigure}
\centering
\begin{subfigure}[b]{0.22\textwidth}
\centering
\includegraphics[width=\textwidth]{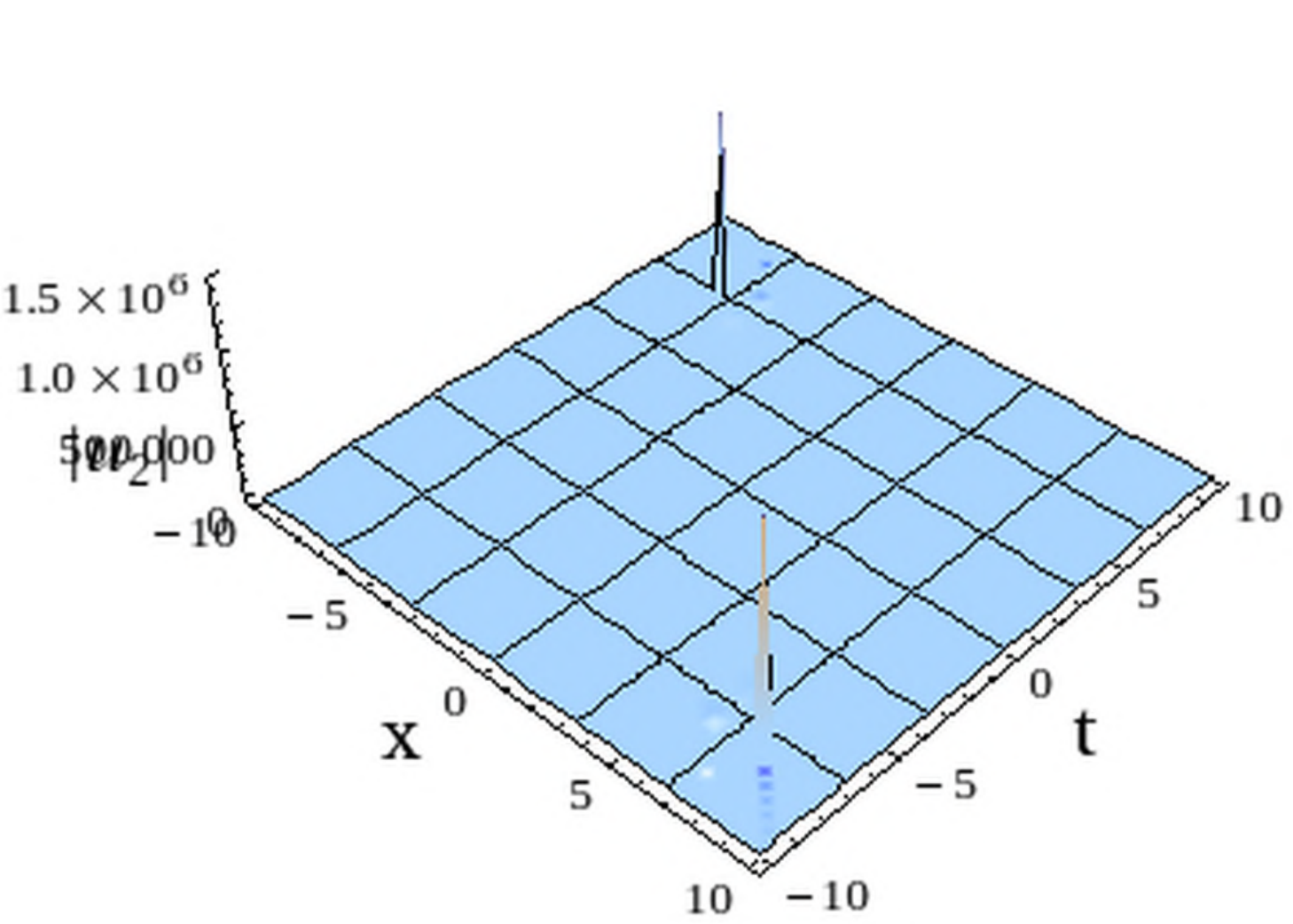}
\caption{}
\label{F1c}
\end{subfigure}
\begin{subfigure}[b]{0.22\textwidth}
\centering
\includegraphics[width=\textwidth]{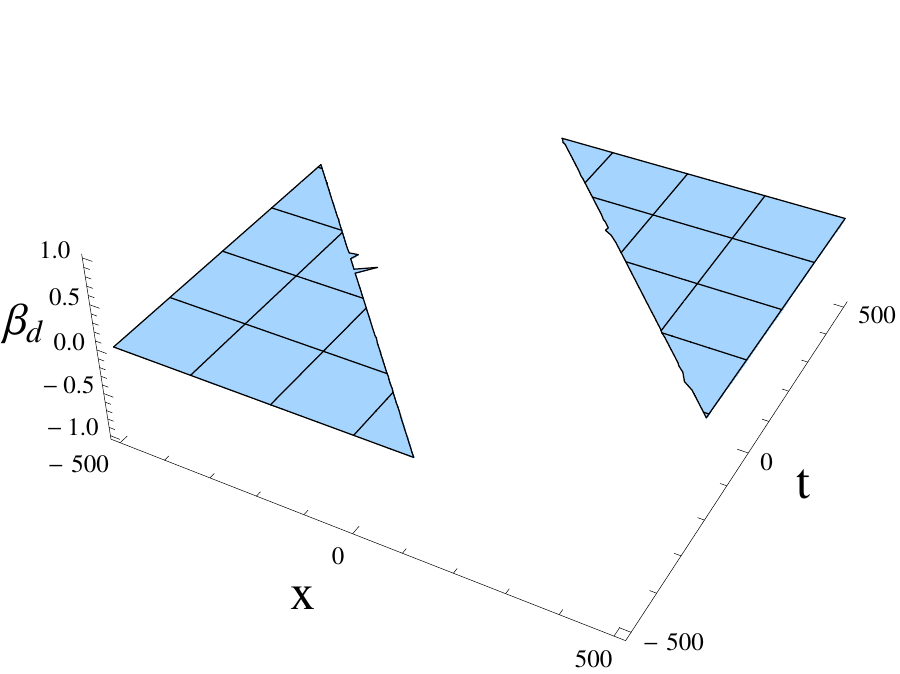}
\caption{}
\label{F1d}
\end{subfigure}
\caption{a) The deformed amplitude $A_d$ is taken as equal to the undeformed 2-soliton amplitude $A_0$ with $a_1=1.1,\,a_2=1,\,\delta=0$. The moduli of the corresponding localized $v_2$ and singular $u_2$ are shown in b) and c) respectively. d) However, the shifted deformed amplitude $\beta_d=B_d-u_1$ does not exist within the localization range of the other amplitudes.}
\end{figure}

However, not every solution to the undeformed AB system, especially the more complicated ones, can serve as an ansatz for the deformed solution $A_d$. For example, extending the treatment to the 2-soliton case by following the general prescription by Gibbons {\it et. al.} \cite{Gibbon79} that pertains to sub-critical sheer in baroclinic fluid, we obtain the expressions as:

\bea
&&A_0^2=4\frac{\partial^2}{\partial x^2}\left[\ln\left\{\det(M)\right\}\right]\nonumber\\
&&\qquad=-4\left(\frac{a_1-a_2}{a_1+a_2}\right)^2\left[\cosh\theta_1\cosh\theta_2-\frac{4a_1a_2}{(a_1+a_2)^2}\cosh^2\frac{\theta_1+\theta_2}{2}\right]^{-2}\nonumber\\
&&\qquad\quad\times\Bigg[\sinh(\theta_1-\theta_2)\left(a_1^2\sinh\theta_1\cosh\theta_2-a_2^2\cosh\theta_1\sinh\theta_2\right)\nonumber\\
&&\qquad\quad-2a_1a_2\left(1+\sinh^2\theta_1\sinh^2\theta_2\right)\Bigg]\nonumber
\eea
and
\bea
&&B_0=-1-\frac{\partial^2}{\partial x\partial t}\left[\ln\left\{\det(M)\right\}\right]\nonumber\\
&&\qquad=
-1-2\left(\frac{a_1-a_2}{a_1+a_2}\right)^2\left[\cosh\theta_1\cosh\theta_2-\frac{4a_1a_2}{(a_1+a_2)^2}\cosh^2\frac{\theta_1+\theta_2}{2}\right]^{-2}\nonumber\\
&&\qquad\times\Bigg[2\sinh\theta_1\sinh\theta_2\left\{1+\cosh(\theta_1+\theta_2)\right\}-\frac{1}{2}\frac{a_1+a_2}{a_1-a_2}\left(1+\cosh2\theta_1\cosh2\theta_2\right)\nonumber\\
&&\qquad\qquad+\frac{a_2^2}{2a_1(a_1-a_2)}\sinh2\theta_1\sinh2\theta_2\Bigg],\label{N03}
\eea
where $a_{1,2}=i\gamma_{1,2}\in\mathbb{R}$, $\theta_{1,2}=a_1x+t/a_1+\delta_{1,2}$ and $M_{ij}=\left(a_i+a_j\right)^{-1}\cosh\frac{1}{2}\left(\theta_i+\theta_j\right)$. On assuming $A_d=A_0$, plotted in Fig. \ref{F1a}, the corresponding moduli of deformation functions $v_2=w_2^*$ and $u_2$ are shown in figures \ref{F1c} and \ref{F1d} respectively. Although $v_2$ is well-localized, $u_2$ displays only singular peaks. This ansatz turns out to be invalid as the deformed and shifted amplitude $B_d-u_1$ turns out to be nonexistent within the localization domain of the others (Fig. \ref{F1d}). A similar situation occurs if we consider the following kink-kink/kink-anti-kink type ansatz for $A_d$,
\bea
&&A_d^{KK/KAK}\nonumber\\
&&=\frac{2}{a}\left[1+\left(\frac{1\mp a^2}{1+a^2}\right)^2\sinh^2\left\{\frac{1\pm a^2}{2a}(x\pm t)\right\}{\rm sech}^2\left\{\frac{1\mp a^2}{2a}(x\mp t)\right\}\right]^{-1}\nonumber\\
&&\quad\times{\rm sech}\left\{\frac{1\mp a^2}{2a}(x\mp t)\right\}\Bigg[\left(1\mp a^2\right)\cosh\left\{\frac{1\pm a^2}{2a}(x\pm t)\right\}\nonumber\\
&&\quad-\frac{\left(1\mp a^2\right)^2}{1\pm a^2}\sinh\left\{\frac{1\pm a^2}{2a}(x\pm t)\right\}\tanh\left\{\frac{1\mp a^2}{2a}(x\mp t)\right\}\Bigg],\label{NH02}\\
&&a\in\mathbb{R}.\nonumber
\eea
as one can verify readily. Such invalid solutions mark the significance of the inhomogeneities that drastically changes the dynamics in general. Interestingly, such kink-anti-kink-type excitations have not been observed in real baroclinic systems. Therefore such non-holonomic inhomogeneous systems could be more suitable for representing physical situations in this regard. 

\begin{figure}
\centering
\begin{subfigure}[b]{0.22\textwidth}
\centering
\includegraphics[width=\textwidth]{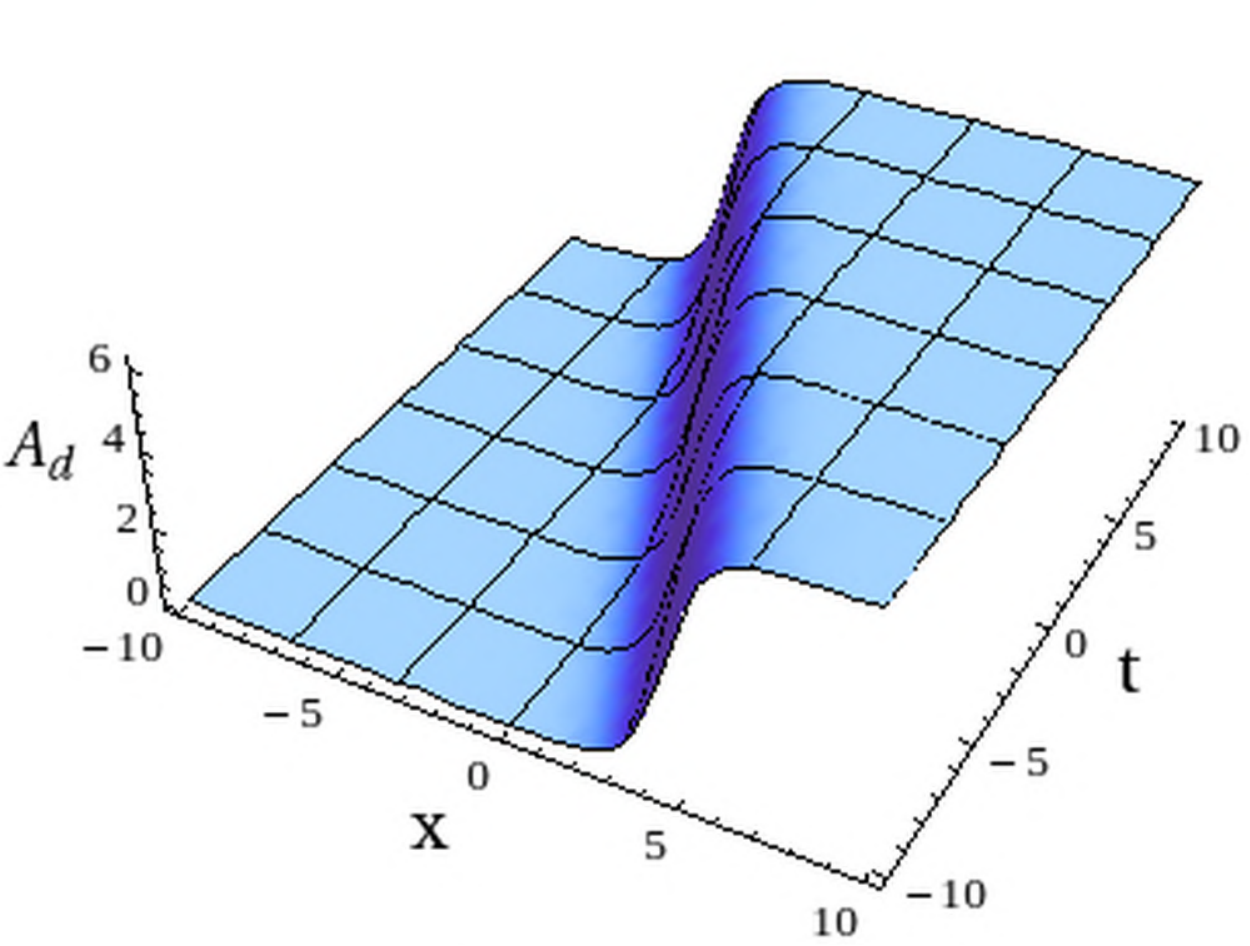}
\caption{}
\label{F3a}
\end{subfigure}
\begin{subfigure}[b]{0.22\textwidth}
\centering
\includegraphics[width=\textwidth]{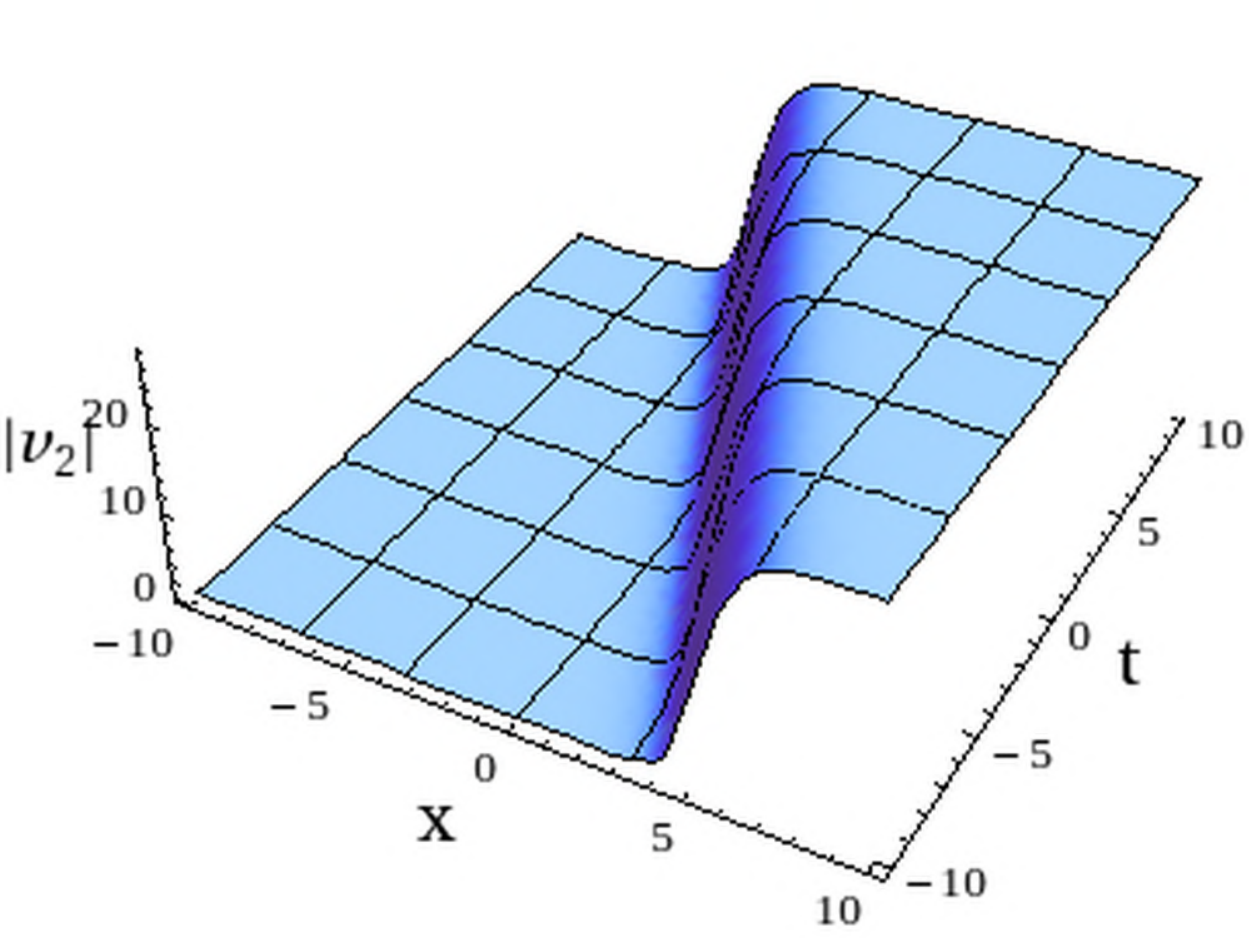}
\caption{}
\label{F3b}
\end{subfigure}
\centering
\begin{subfigure}[b]{0.22\textwidth}
\centering
\includegraphics[width=\textwidth]{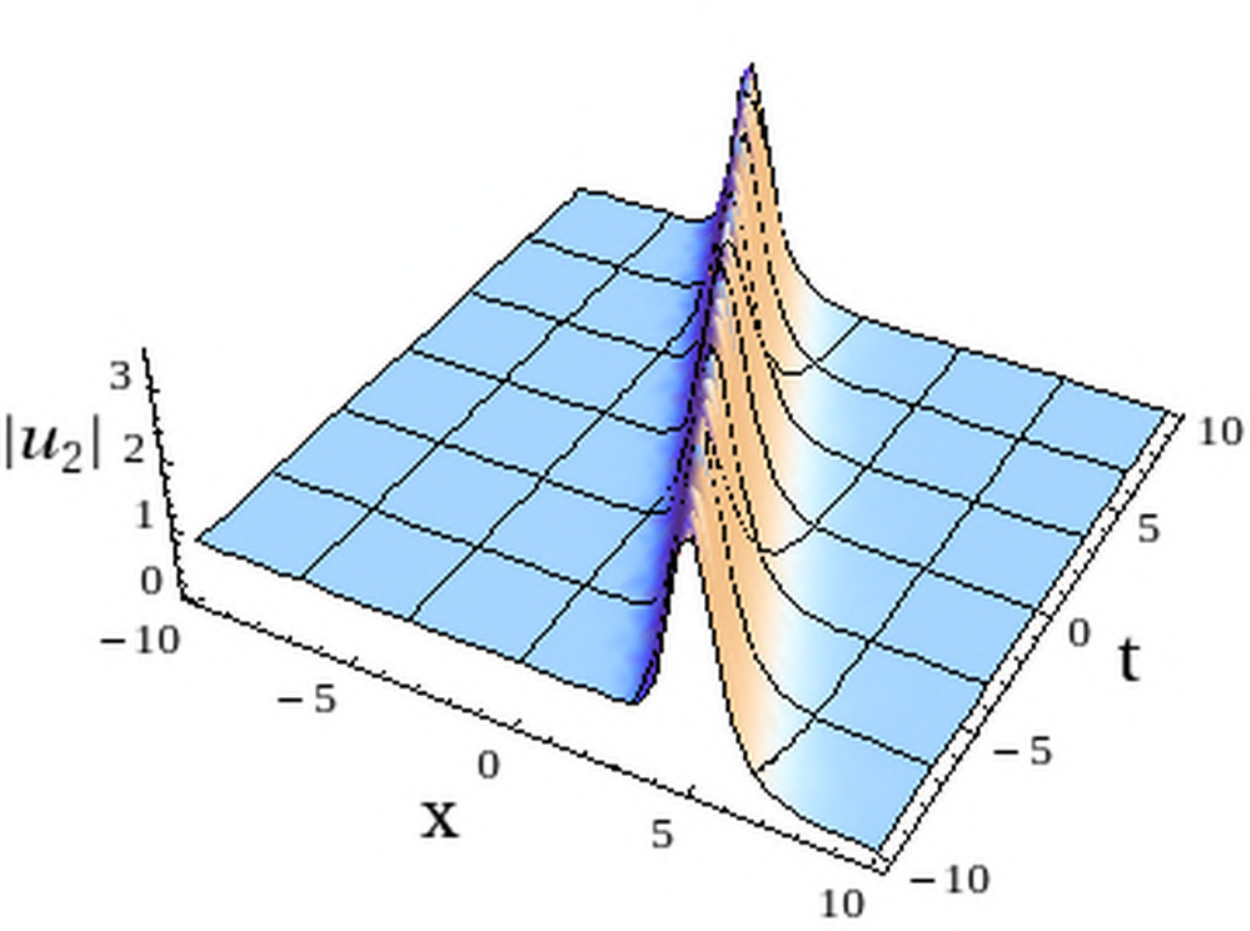}
\caption{}
\label{F3c}
\end{subfigure}
\begin{subfigure}[b]{0.22\textwidth}
\centering
\includegraphics[width=\textwidth]{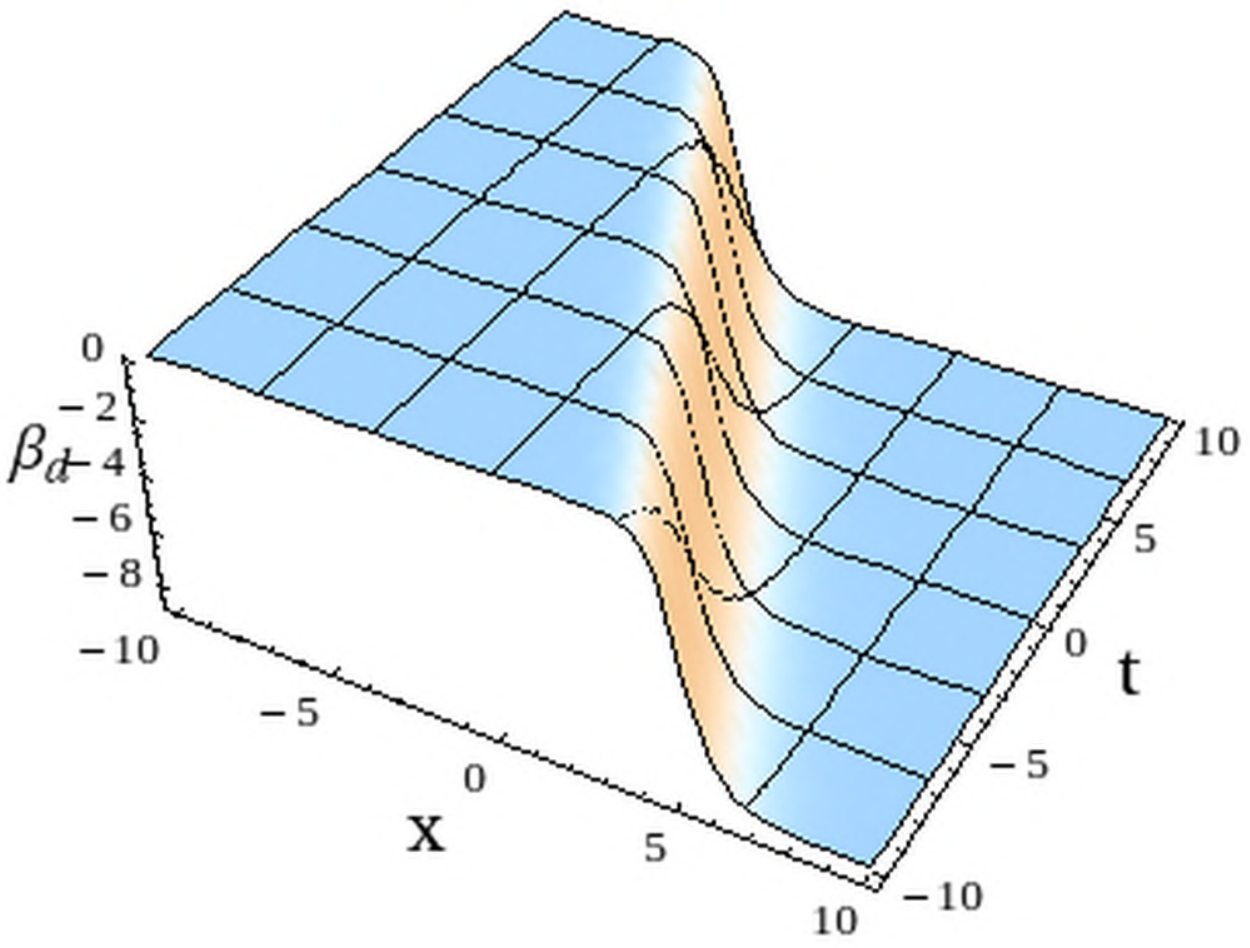}
\caption{}
\label{F3d}
\end{subfigure}
\caption{a) A kink-type ansatz is adopted for the deformed amplitude $A_d$ with $a_1=1.5,\,\delta=0$. The moduli of the corresponding deformation functions $v_2$ and $u_2$ are subsequently well-behaved, as shown in b) and c) respectively. d) Finally, the shifted deformed amplitude $\beta_d=B_d-u_1$ exists and fairly localized in order to validate the present ansatz, marking integrability.}
\end{figure}

The most general and obvious approach to obtain a deformed solution is to start with particular deformation functions $\left(u_2,\,v_2\right)$ and then to solve for $\left(A_d,\,B_d-u_1\right)$ using Eq.s \ref{NN06}. However it might not be easy to achieve that analytically given the coupled nonlinear nature of the deformed pair of equations. This initial study of non-holonomically deformed AB system is content with obtaining a few localized deformed solutions via judicious ansatz thereby ensuring physically realizable states. As the next solution we consider an ansatz for $A_d$ which is {\it not} a solution to the undeformed system, namely, a kink-type structure:

\be
A_d=4\arctan\left[\exp\left(ax+\frac{t}{a}+\delta\right)\right],\quad a,\delta\in\mathbb{R},\label{NH03}
\ee
shown in Fig. \ref{F3a}. The motivation behind the above ansatz is the known kink-type solutions in both sine-Gordon and quasi-sine-Gordon systems \cite{A1,A7}. The corresponding deformation functions,

\bea
&&v_2=-2i\left[\frac{2e^{3\theta}}{\left(1+e^{2\theta}\right)^2}-\frac{e^\theta}{1+e^{2\theta}}-\frac{8}{a^2}\left\{\arctan\left(e^\theta\right)\right\}^3\right]\equiv w_2^*\quad{\rm and}\nonumber\\
&&u_2=\frac{i}{2\arctan\left(e^\theta\right)}\Bigg[-\frac{8a e^{5\theta}}{\left(1+e^{2\theta}\right)^3}+\frac{8 e^{3\theta}}{\left(1+e^{2\theta}\right)^2}-\frac{a e^\theta}{1+e^{2\theta}}\nonumber\\
&&\qquad-\frac{24 e^\theta}{a\left(1+e^{2\theta}\right)}\left\{\arctan\left(e^\theta\right)\right\}^2\Bigg],\quad{\rm where}\label{NH04}\\
&&\theta=at+\frac{t}{a}+\delta,\nonumber
\eea
are both localized and well-behaved (Fig.s \ref{F3b} and \ref{F3c} respectively). Eventually, the other deformed and shifted amplitude turns out to be of anti-kink type:

\be
B_d-u_1=\frac{8}{a^2}\left[\arctan\left(e^\theta\right)\right]^2,\label{NH05}
\ee
as plotted in Fig. \ref{F3d}. Therefore the present ansatz is consistent with the NHD. Unique to the deformed system, such solutions may correspond to additional topological or boundary conditions on the AB system. Evidently kink-type structures can occur in case of long neutral modes in a single atmospheric layer away from the baroclinic instability \cite{Muller}. It may be possible that a non-holonomic AB generalization can model such a system. Further, kink or anti-kink solutions are standard solutions of the sine-Gordon equation \cite{Perring}.

As the deformed AB system is a new integrable system with sources subjected to additional constraints, there can be more deformed solutions obtained in the above manner, without having any counterpart in the undeformed sector. Although less general than solving directly for a given set $\left(u_2,\,v_2\right)$, such ansatz provide an elegant way of finding physically interesting solutions for a system which is just a perturbation away from the standard AB system. Such solutions can possibly represent some actual situations in atmospheric and geophysical systems. 

\section{QID of the AB system}
Real physical systems incorporate irregularities and/or deformities which do not get translated consistently when approximated by integrable field theories possessing an infinity of conserved quantities. These conserved quantities are called charges in the N\"other sense \cite{Noether}, pertaining to complete solvability of the continuum system with infinite degrees of freedom. Charges of an integrable system can directly be obtained from the zero curvature condition itself in 1+1 dimensions. An infinity of conserved charges also imply infinite symmetries of the field theory which, however, cannot be consistent with the aforementioned deformities of the actual system. Therefore a more accurate representation can be a slight deformation of an integrable model with a subset of the charges being locally non-conserved that can correspond to the said deformities. Of particular interest is the situation when these non-conserved charges regain conservation asymptotically given certain symmetries of {\it particular} solutions. As a result the system can support localized structures like solitons in absence of local interactions among them \cite{A1,A7}. Additionally, in case the deviation from integrability is sufficiently moderate then standard techniques such as Lax pair or inverse scattering \cite{A1} and pseudo-potential approach \cite{RSG,RKdV} are still applicable, although the system may not have infinite conserved quantities which are in involution. Realistic systems that can support robust collective modes are more accurately represented by such almost-integrable systems.

The QID of integrable models particularly satisfy such requirements, as seen in the case of sine-Gordon \cite{A2,A1,A7} and NLS \cite{A3} systems in the recent past, along with their generalizations \cite{A4,AGM1}. They all possess soliton-like configurations similar to solitons of known integrable models \cite{A5,A6}, such as the baby Skyrme models with many potentials and the Ward modified chiral models \cite{A7}. These systems also display definite space-time parity properties preserved over the quasi-deformation \cite{A3,A1} which has been speculated as a necessary condition for quasi-integrability \cite{tB}. The observation is that the rate of change of anomalous charges are odd in parity about some point in space-time and thus vanish upon integration between $t=\pm\infty$.

The QID of any integrable system is of importance not only due to possible physical applications but also to analyze localized solution sectors in a less restrictive scenario. Herein we quasi-deform the AB system which had not been achieved in preceding literature to the best of our knowledge. The most common prescription for QID is through the zero curvature formalism wherein the non-linear (potential) part of the dynamical equation is suitably deformed by aptly altering the corresponding terms in the {\it temporal} Lax component \cite{A3,A1}. Modification to the temporal Lax component preserves the scattering data though the time-evolution is effected, as mentioned in the NHD case. The resultant {\it non-zero} curvature yields an {\it anomaly} term, followed by a standard Abelianization procedure wherein the spatial Lax component is gauge-rotated using the inherent loop-algebra of the system to obtain linear equations of dynamic coefficients serving as continuity equations \cite{A3,A1}. Eventually, an infinite number of quasi- or locally non-conserved charges are obtained which are directly dependent on the aforementioned anomaly. Given these charges are asymptotically conserved subjected to particular parity properties of the solutions, the system is deemed to be quasi-integrable \cite{A3,tB,A2,A1,A7}. Very recently it has been shown that an extended infinite tower of quasi-conserved charges for deformed sine-Gordon \cite{RSG} and particular deformations of KdV \cite{RKdV} systems, can be obtained through the Riccati-type pseudo-potential approach \cite{RTSP1,RTSP2} deeming them to be quasi-integrable. This approach could have the advantage of directly calculating quasi-conservation laws over the zero-curvature approach.

However, in the present work, we employ the zero curvature approach owing to its clear algebraic demonstration of the quasi-deformation procedure and the subsequent symmetry (parity) conditions. In order to obtain the anomalous curvature for a dual system like AB, however, the aforementioned potential term is not apparent (Eq. \ref{1}). To resolve this issue the fact that when $A_0$ is real the AB system can lead to the sine-Gordon equation under the following identifications \cite{Gibbon}:

\be
A_0=\psi_x,\quad B_0=\cos\psi,\label{21}
\ee
is utilized. In terms of $\psi$ simultaneity of {\it both} the AB equations forces the sine-Gordon equation $\psi_{xt}=\sin\psi$ to be satisfied with $(x,t)$ identified as suitable light-cone coordinates. Although $A_0$ is complex in general, as both the systems are integrable a viable quasi-deformation scheme for the AB system should be analogous to that for the sine-Gordon case. Following the quasi-deformation of the sine-Gordon equation \cite{A2,A1} wherein the nonlinear or `potential' term $V_{SG}(\psi)=\cos\psi$ was deformed the obvious choice for quasi-deformation in the AB case has to be the function $B_0$. Further, by substituting $B_0$ from the first of Eq.s \ref{1} in the second,

\be
A_{0,xt}=-\frac{1}{2}A_0\partial^{-1}_x\left(\vert A_0\vert^2\right)_t,\label{22}
\ee
marks $B_0$ as the nonlinear term in the evolution of $A_0$ and thereby equivalently is the `potential' of the theory.

Therefore a suitable QID should correspond to deforming the $B_0$-dependent part in the temporal Lax component $M$ in Eq. \ref{2} in analogy with the sine-Gordon system \cite{A1}. To our comfort (to be explicated later), the $B_0$-dependent contribution to the Lax pair exclusively belongs to the Kernel subspace of the corresponding $sl(2)$ loop algebra in $M$ which is crucial condition for obtaining a large class of quasi-deformed systems \cite{A3,AGM1}. Following this analogy, $B_0$ itself corresponds to the sine-Gordon potential, $V_{SG}(\psi)=\frac{1}{16}\left(1-\cos\psi\right)$ and thus, the quasi-sine-Gordon system should correspond to an exclusive deformation $B_0\to B$ when $A_0\to A$ is real, which amounts to \cite{A1}:

\be
V_{SG}(\psi)\to \frac{2}{(2+\varepsilon)^2}\tan^2\frac{\psi}{4}\left(1-\left|\sin\frac{\psi}{4}\right|^{2+\varepsilon}\right)^2\equiv V_{SG}^\varepsilon,\quad\psi=\partial_x^{-1}A,\label{24}
\ee
where $\varepsilon$ being the parameter of deformation. In terms of the AB system parameters this deformation can be expressed as,

\be
B_0\to B=1-\frac{32}{(2+\varepsilon)^2}\frac{\sqrt{2}-\sqrt{1+\tilde{B}}}{\sqrt{2}+\sqrt{1+\tilde{B}}}\left[1-\left\{\frac{1}{2\sqrt{2}}\left(\sqrt{2}-\sqrt{1+\tilde{B}}\right)\right\}^{1+\frac{\varepsilon}{2}}\right]^2.\label{23}
\ee
The somewhat `intermediate' deformed function $\tilde{B}=\cos\Psi$, introduced for convenience of the deformation, can be visualized in terms of a deformed sine-Gordon system with $\Psi$ satisfying $A=\Psi_x$. 

For $A_0$ being complex, the simplest generalization to the sine-Gordon mapping implies:

\be
A_0=\psi_x e^{i\phi},\quad B_0=\cos\psi:\quad \psi,\phi\in\mathbb{R}.\label{N1}
\ee
As $B_0$ and thereby $\tilde{B}$ and $B$ still being real through Eq. \ref{24}, except for the updated definition $\psi= \partial^{-1}_x\left(e^{-i\phi} A_0\right)$ which still is a real solution of the sine-Gordon equation. Subsequently only the potential function $B_0$ is deformed under QID but not $\psi$ itself. Thus we get a QID scheme for the case when the AB-system is complex, still maintaining the analogy with the corresponding sine-Gordon system. We proceed with the above definition from hereon, although there is nothing to suggest that this is the only way to quasi-deform the AB system. The zero curvature condition now takes the form:

\be
F_{tx}=\frac{1}{8i\lambda}\left[2B_x+\left(\vert A\vert^2\right)_t\right]\sigma_3+\frac{1}{4i\lambda}\left[-A_{xt}+B A\right]\sigma_++\frac{1}{4i\lambda}\left[-A^*_{xt}+B A^*\right]\sigma_-.\label{8}
\ee
Starting from Eq.s \ref{2} the deformation $B_0\to B$ changes the system to one marked by the deformed solutions of $A_{xt}=AB$. This deformed `on-shell' condition would imply, in general, that $2B_x+\left(\vert A\vert^2\right)_t\neq 0$. Therefore the corresponding curvature does not vanish: $F_{tx}={\mathcal X}\sigma_3/\lambda\neq 0$ and the system is no longer integrable. The non-vanishing contribution: 

\be
{\mathcal X}:=-\frac{i}{8}\left[2B_x+\left(\vert A\vert^2\right)_t\right]\neq 0.\label{28}
\ee
is termed as the {\it anomaly} pertaining to QID. Obviously, the normalization condition: $\vert A_{0,t}\vert^2+B_0^2=1$ is no longer supported by the quasi-deformed system more so as the very integrability of the system is disturbed. 

\subsection{The Abelianization and quasi-conserved charges}
In order to obtain the quasi-conserved charges of the deformed AB system the corresponding $sl(2)$ loop algebra is to be utilized \cite{A1} which is constructed from the basic $SU(2)$ structure as \cite{AGM1}:

\bea
&&\left[b^n,\,F^m_{1,2}\right]=2F^{m+n}_{2,1},\quad \left[F^n_1,\,F^m_2\right]=\kappa b^{m+n}\quad{\rm where},\nonumber\\
&& b^n=\lambda^n\sigma_3,\quad F^n_{1,2}=\frac{\lambda^n}{\sqrt{2}}\left(\kappa\sigma_+\mp\sigma_-\right),\label{11}
\eea
where $\kappa$ is a number. The goal is to linearize the system in order to obtain quasi-continuity equations that directly follow from the flatness condition. As a result we obtain quantities, identified as charges, whose time variation would have vanished in the absence of the anomaly term. To this end a gauge transformation is to be carried out which is independent of the Kernal subspace of the $sl(2)$ algebra, characterized by the generators $b^n$, and spanned in its Image as \cite{A3}:

\be
g=\exp\sum_{n=1}^\infty J_{-n},\quad J_{-n}=a_1^{-n}F_1^{-n}+a_2^{-n}F_2^{-n};\quad a_{1,2}\in\mathbb{C}.\label{12}
\ee
The gauge transformation is chosen such that one of the Lax components exclusively belong to the Kernel subspace which will Abelianize the system by leading to linear relations (continuity equations) among gauge-rotated coefficients in that subspace. Conventionally the spatial Lax component:
\be
L=-ib^1+\frac{1}{2\sqrt{2}}\left(\frac{A}{\kappa}+A^*\right)F_1^0+\frac{1}{2\sqrt{2}}\left(\frac{A}{\kappa}-A^*\right)F_2^0,\label{NR1}
\ee
re-expressed in terms of the $sl(2)$ generators, is chosen as its time derivative appears in the zero curvature condition. There is a semi-simple element of the $sl(2)$ algebra ($-ib^1$) in $L$, which is constant and has a different grade than the rest of the terms in $L$, and thus can split the algebra into the corresponding Kernel and Image subspaces. Accordingly, the element $g$ of the gauge group is chosen to contain only negative spectral powers ($n\ge1$). As a result, under the gauge transformation $L\to\Bar{L}=gLg^{-1}+g_xg^{-1}$, the rotated operator $\bar{L}$ is exclusive to the Kernel of $sl(2)$:

\be
\Bar{L}=\sum_n\beta_L^{-n}b^{-n},\label{14a}
\ee
since the rotated expansion coefficients of $L$ do not get mixed. As a result certain consistency conditions are imposed determining the expansion coefficients $a_{1,2}^{-n}$ in $g$, as shown in Eq.s \ref{16} of the appendix, owing to the constant semi-simple element in $L$. The surviving coefficients $\beta_L^{-n}$ are then obtained in Eqs. \ref{17}, which are crucial for linearizing the system and thereby leads to the notion of charges. Subsequently, the temporal Lax component gauge-transforms as,

\bea
&&M=i\frac{B}{4}b^{-1}-i\frac{1}{4\sqrt{2}}A_{-,t}F_1^{-1}-i\frac{1}{4\sqrt{2}\kappa}A_{+,t}F_2^{-1}\nonumber\\
&&\quad\to\Bar{M}=gMg^{-1}+g_tg^{-1}=\sum_n\left[\beta_M^{-n}b^{-n}+\alpha_1^{-n}F_1^{-n}+\alpha_2^{-n}F_2^{-n}\right],\label{19}
\eea
spanning whole of the $sl(2)$ group space in general. The $B$-depended contribution to $M$ is exclusive to the $sl(2)$ Kernel sub-space that mirrors the location of the deformed potential in case of the quasi-sine-Gordon system \cite{A1,A7}, further akin to the quasi-NLS \cite{A3} system and its extensions \cite{AGM1}. As iterated earlier, such an algebraic resemblance assures consistent results upon the standard Abelianization procedure of quasi-deformation, more so as the AB system is a generalization of the sine-Gordon system. We evaluate the coefficients of the gauge-rotated temporal Lax component in Eq.s \ref{21a} of the appendix. 

The gauge-rotated quasi-deformed curvature also transforms under this gauge rotation as: 

\bea
&&F_{tx}={\mathcal X}b^{-1}\to\Bar{F}_{tx}=gF_{tx}g^{-1}\equiv{\mathcal X}gb^{-1}g^{-1},\nonumber\\
&&\qquad:={\mathcal X}\sum_n\left(f_0^{-n}b^{-n}+f_1^{-n}F_1^{-n}+f_2^{-n}F_2^{-n}\right).\label{22aa}
\eea
A few of the corresponding expansion coefficients are listed in Eq.s \ref{24a} in the appendix. Consistency demands that the above expression of the rotated curvature, obtained through direct gauge transformation, must be the same as that obtained in terms of the transformed Lax pair $\left(\Bar{L},\Bar{M}\right)$ which is,

\bea
&&\Bar{F}_{tx}=\Bar{L}_t-\Bar{M}_x+\left[\Bar{L},\,\Bar{M}\right]\nonumber\\
&&\qquad=\sum_n\left[\left(\beta^{-n}_{L,\,t}-\beta^{-n}_{M,\,x}\right)b^{-n}-\alpha^{-n}_{1,\,x}F_1^{-n}-\alpha^{-n}_{2,\,x}F_2^{-n}\right]\nonumber\\
&&\qquad\qquad+2\sum_{m,n}\beta_L^{-n}\left(\alpha_1^{-m}F_2^{-m-n}+\alpha_2^{-m}F_1^{-m-n}\right).\label{25}
\eea
By equating the coefficients of the rotated curvature obtained in both ways we finally obtain two sets of linearized equations:

\bea
&&\beta_{L,\,t}^{-n}-\beta_{M,\,x}^{-n}={\mathcal X}f_0^{-n}\quad\text{and}\nonumber\\
&&-\alpha_{(1,2)\,x}^{-n}+2\sum_m\beta_L^{-n}\alpha_{2,1}^{-m-n}={\mathcal X}f_{1,2}^{-n}.\label{26a}
\eea
The first set of the above equation represents the Abelianization of the system and thereby constitutes an infinite set of continuity equations. Naturally, the definition of an infinite set of charges follow as,

\be
Q^{-n}:=\int_x\beta_L^{-n},\label{NR3}
\ee
whose time-evolution solely depend on the QID anomaly ${\mathcal X}$:

\be
\frac{dQ^{-n}}{dt}=\int_x\beta_{L,t}^{-n}=\int_x\left(\beta_{M,\,x}^{-n}+{\mathcal X}f_0^{-n}\right)\equiv\int_x{\mathcal X}f_0^{-n},\label{27}
\ee
where the last step considers suitable boundary conditions. However, the notion of boundary here is non-trivial in view of quasi-integrability as, for a physical system, the boundary is usually finite. In that case, we assume the solutions ($A,\,B$) to be well-localized so that they can be assumed to vanish at sufficiently large boundaries. This assumption is supported by numerous observations of solitonic excitations in real continuous systems \cite{Dodd,Rajaraman}. Moreover, given the definite parity properties crucial for quasi-integrability, the boundaries are assumed to be symmetric about the corresponding point of reflection\cite{A2,A1,A7}. 

A few of the lowest order charges and their conservation can be checked right away. For $n=1$ we have,

\be
\frac{dQ^{-1}}{dt}=\int_x{\mathcal X}\equiv-\frac{i}{8}\frac{d}{dt}\int_x\vert A\vert^2,\label{N02}
\ee
with the deformed potential $B$ being asymptotically well-behaved. The RHS above is nothing but the time-evolution of {\it total density} of the system with solution $A$. As expected, a physical system has a well-localized solution profile and so a conserved $Q^{-1}$, which we will find out to be the case. Such an assertion is true for particular localized solutions of the deformed system which we will obtain later. However, the quasi-conservation of the system is not assured yet since asymptotic conservation of all the charges is yet to be established given definite space-time parity of the solution. Presently, a trivial case is that of $n=2$; since $f_0^{-2}=0$ the corresponding charge $Q^{-2}$ is conserved identically. Such occurrences of local conservation in quasi-deformed system is known \cite{A2,A1,A7}, which may correspond to particular residual robust symmetries of the system. The next couple of charge evolution relations are:

\bea
&&\frac{dQ^{-3}}{dt}=-\frac{\kappa}{32}\int_x{\mathcal X}\left(A_+^2-A_-^2\right)=-\frac{1}{8}\int_x{\mathcal X}\vert A\vert^2,\nonumber\\
&&\frac{dQ^{-4}}{dt}=i\frac{\kappa}{32}\int_x{\mathcal X}\left(A_{+,x}A_--A_{-,x}A_+\right)=-\frac{i}{16}\int_x{\mathcal X}\left(A_xA^*-A^*_xA\right),\label{NN03}
\eea
which are non-vanishing in general. Here $A_\pm=\frac{A}{\kappa}\pm A*$.

As has been stated, quasi-conservation essentially amounts to asymptotic conservation of the charges, including the situation when the localized solutions with definite parity are well-separated \cite{tB}. Therefore in order to explicitly check for quasi-conservation one needs to obtain particular solutions ($A,B$) of the the deformed system. However, even before trying to obtain those solutions, one can exploit the group-algebraic nature of the Abelianization process ensuring quasi-integrability of the system, namely the behavior of the different operators under space-time parity and $sl(2)$ automorphism which we will discuss next.

\subsection{General Quasi-integrability Structure}
The known quasi-integrable systems are found to display definite properties under the $\mathbb{Z}_2$ transformation of the $sl(2)$ loop algebra which essentially ensure fixed values of $Q^{-n}$ at the space-time boundaries \cite{A3,A1,AGM1}. The $\mathbb{Z}_2$ transformation is a product of the order 2 automorphism of the $sl(2)$ loop algebra:

\be
\Sigma\left(b^n\right)=-b^n,\quad\Sigma\left(F_{1,2}^n\right)=\mp F_{1,2}^n
\ee
and space-time parity:

\be
{\mathcal P}:\quad \left(\tilde{x},\,\tilde{t}\right)\to\left(-\tilde{x},\,-\tilde{t}\right),\quad \tilde{x}=x-x_0,~\tilde{t}=t-t_0,
\ee
about some arbitrary point $\left(x_0,\,t_0\right)$ which can very well chosen to be the origin. Trivially, the order-2 automorphism and parity  operations commute and the $\mathbb{Z}_2$ operation is labelled as $\Omega={\mathcal P}\Sigma$. We will demonstrate, as is the case for other known quasi-conserved systems, that definite $\mathbb{Z}_2$ symmetry of the spatial Lax component $L$ leads to definite parity of the anomaly and the expansion coefficients over the Abelianizing rotation which in turn ensures quasi-conservation.

One immediately observes (Eq. \ref{NR1}) that $L$ can have definite $\mathbb{Z}_2$ behavior only if $A_\pm$ have definite parities. Given the semi-simple element in $L$ that splits the algebra ($b^1$) has a constant coefficient the only possibility is that,

\be
\Omega\left(L\right)=-L.
\ee
This mandates that the fields $A_\pm$ to be parity-even/odd respectively. However, for $A=A_R+iA_I,~A_{R,I}\in\mathbb{R}$,

\be
A_\pm=\left(\frac{1}{\kappa}\pm 1\right)A_R+i\left(\frac{1}{\kappa}\mp 1\right)A_I.
\ee
Thus $A_\pm$ cannot have the desired parity properties unless,

\begin{enumerate}
     \item either $\kappa=1$, ${\mathcal P}\left(A_R\right)=A_R$ and ${\mathcal P}\left(A_I\right)=-A_I$,
    \item or $\kappa=-1$, ${\mathcal P}\left(A_R\right)=-A_R$ and ${\mathcal P}\left(A_I\right)=A_I$.
\end{enumerate}
Consistency of the choice ${\mathcal P}\left(A_\pm\right)=\pm A_\pm$ is demonstrated by the fact that it immediately ensures definite parity of the expansion coefficients of the gauge element $g$:

\bea
{\mathcal P}\left(a_{1,2}^{-n}\right)=\mp a_{1,2}^{-n},
\eea
rendering $\Omega\left(g\right)=g$. Further the expansion coefficients of the rotated component $\bar{L}$ in the Kernel satisfy ${\mathcal P}\left(\beta_L^{-n}\right)=\beta_L^{-n}$ leading to,

\be
\Omega\left(\bar{L}\right)=-\bar{L}.
\ee
Thus the definite $\mathbb{Z}_2$ behavior of the spatial component is preserved over the gauge transformation.

As a general demonstration of this preservation over gauge rotation let us consider a general gauge transformation through,

\be
\mathfrak{g}=\exp\sum_{n=1}^\infty\mathfrak{G}^{-n},
\ee
where $\mathfrak{G}^n$ is any linear combination of $F_{1,2}^{-n}$ confined to the Image. The rotated spatial component $\mathfrak{L}=\mathfrak{g}L\mathfrak{g}^{-1}+\mathfrak{g}_x\mathfrak{g}^{-1}$ will now have an expansion $\mathfrak{L}=\sum_n\mathfrak{L}_{n}$ in the Kernel. An order-by-order comparison in spectral powers following the well-known Baker–Campbell–Hausdorff (BCH) expansion yields,

\bea
&&\mathfrak{L}_{1}=L_1,\nonumber\\
&&\mathfrak{L}_{0}=L_0+\left[\mathfrak{G}^{-1},\,L_1\right],\nonumber\\
&&\mathfrak{L}_{-1}=\left[\mathfrak{G}^{-1},\,L_0\right]+\left[\mathfrak{G}^{-2},\,L_1\right]+\frac{1}{2!}\left[\mathfrak{G}^{-1},\left[\mathfrak{G}^{-1},\,L_1\right]\right]+\mathfrak{G}^{-1}_x,\nonumber\\
&&\mathfrak{L}_{-2}=\left[\mathfrak{G}^{-2},\,L_0\right]+\frac{1}{2!}\left[\mathfrak{G}^{-1},\left[\mathfrak{G}^{-1},\,L_0\right]\right]+\left[\mathfrak{G}^{-3},\,L_1\right]\nonumber\\
&&\qquad\quad+\frac{1}{3!}\left[\mathfrak{G}^{-1},[\left[\mathfrak{G}^{-1},\left[\mathfrak{G}^{-1},\,L_1\right]\right]\right]+\frac{1}{2!}\Big(\left[\mathfrak{G}^{-2},\left[\mathfrak{G}^{-1},\,L_1\right]\right]\nonumber\\
&&\qquad\quad+\left[\mathfrak{G}^{-1},\left[\mathfrak{G}^{-2},\,L_1\right]\right]\Big)+\mathfrak{G}^{-2}_x+\frac{1}{2!}\left[\mathfrak{G}^{-1},\,\mathfrak{G}^{-1}_x\right],\nonumber\\
&&\vdots\label{NR2}\\
&&{\rm where},\nonumber\\
&&L=L_0+L_1,\quad L_0=\frac{1}{2\sqrt{2}}A_+F_1^0+\frac{1}{2\sqrt{2}}A_-F_1^0,~L_1=-ib^1.\nonumber
\eea
From the first equation above, $\Omega\left(\mathfrak{L}_{1}\right)=-\mathfrak{L}_{1}$. The second one implies,

\be
\Omega\left(\mathfrak{L}_0\right)=-L_0-\left[\Omega\left(\mathfrak{G}^{-1}\right),\,L_1\right].
\ee
On adding it back to the second of Eq.s \ref{NR2},

\be
\left(1+\Omega\right)\left(\mathfrak{L}_0\right)=\left[\left(1-\Omega\right)\left(\mathfrak{G}^{-1}\right),\,L_1\right].
\ee
The LHS above belongs to the Kernel whereas the RHS is limited to the Image of $SL(2)$; hence both must vanish, leading to,

\be
\Omega\left(\mathfrak{L}_0\right)=-\mathfrak{L}_0\quad{\rm and}\quad \Omega\left(\mathfrak{G}^{-1}\right)=\mathfrak{G}^{-1}.
\ee
Similarly from the subsequent equation one successively obtains,

\be
\Omega\left(\mathfrak{L}_{-n}\right)=-\mathfrak{L}_{-n}\quad{\rm and}\quad \Omega\left(\mathfrak{G}^{-n-1}\right)=\mathfrak{G}^{-n-1}\quad\forall~n.
\ee
Finally the culminated result is $\Omega\left(\mathfrak{g}\right)=\mathfrak{g}$ and thus $\Omega\left(\mathfrak{L}\right)=-\mathfrak{L}$ as obtained earlier for the particular case of $\mathfrak{G}=g$. Thus for a gauge element $\mathfrak{G}$ exclusive to the Image subspace the definite $\mathbb{Z}_2$ automorphism of the spatial component is preserved over the Abelianizing gauge rotation. However the definite parities of $A_\pm$ required for definite $\mathbb{Z}_2$ automorphism of $L$ to begin with is a must. In fact the relatively trivial choices to keep $\Omega\left(L\right)=-L$, namely,
\begin{itemize}
    \item[i)] either $\kappa=1$, ${\mathcal P}\left(A_R\right)=A_R$, $A_I=0$
    \item[ii)] or $\kappa=1$, ${\mathcal P}\left(A_I\right)=A_I$, $A_R=0$ 
\end{itemize}
both imply ${\mathcal P}\left(A_+\right)=A_+$ and $A_-=0$. In the undeformed case, the first choice reduces the AB system to the sine-Gordon system whereas the second to the sinh-Gordon system both of which are well-known to be quasi-deformable \cite{A1,A7}. Since the definite $\mathbb{Z}_2$ automorphism of the Lax component is a mainstay of the known quasi-deformable systems \cite{A3,tB,A1,AGM1,OwnKdV}, the necessary condition(s) for obtaining the same for AB system leading to known quasi-deformable systems strongly supports quasi-deformability of the AB system. One can recheck that the expansion coefficients of both $g$ and $\bar{L}$ either demonstrate correct parities for these {\it reduced} choices or vanish altogether.

The temporal Lax component $M$ spans both Kernel and Image subspaces. Still it is straight-forward to see that $M$ also maintains definite $\mathbb{Z}_2$ automorphism after Abelianization. However, apart from the prevailing choice ${\mathcal P}\left(A_\pm\right)=\pm A_\pm$, definite $\mathbb{Z}_2$ automorphism of $M$ further requires ${\mathcal P}\left(B\right)=B$. Using the individual parity properties one can verify that $\Omega\left(M\right)=-M$ and eventually, from the corresponding expansion coefficients derived above, that $\Omega\left(\bar{M}\right)=-\bar{M}$. Equivalently, one can also consider the effect of a general Abelianizing rotation through $\mathfrak{G}$ and use the known $\mathbb{Z}_2$ automorphism properties to obtain the same result. 

However, the expansion coefficients that directly appear in the expression of the charge are from $\bar{L}$ and the rotated curvature $\bar{F}_{tx}$. From Eq. \ref{NR3}, since $\{\beta_L^{-n}\}$ are parity-even, the charges $Q^{-n}$ are non-trivially finite and time-dependent\footnote{Unless of course $\beta_L^{-n}=0$ itself which is the case for $n=0$.}. The time-variation of $\{\beta_L^{-n}\}$ which is of prime importance for conservation, however, depends on the expansion coefficients $\{f_0^{n}\}$ of the rotated curvature in the Kernel. From our explicit expressions and previous results it can be readily seen that ${\mathcal P}\left(f_0^{-n}\right)=f_0^{-n}$. To show it more directly we utilize the killing form of the $sl(2)$ loop algebra:

\be
\mathfrak{K}(*)=\frac{1}{i2\pi}\oint\frac{d\lambda}{\lambda}{\rm Tr}(*),
\ee
where the trace is taken in the $SU(2)$ space. For the present choice of $sl(2)$ generators,

\be
\mathfrak{K}(b^nb^m)=\frac{1}{2}\delta_{n+m,0},\quad \mathfrak{K}(F_{1,2}^nF_{1,2}^m)=\mp\kappa\delta_{n+m,0},
\ee
with all other combinations vanishing. Then, on utilizing Eq. \ref{22a}, one has,

\be
f_0^{-n}=2\kappa\left(gb^{-1}g^{-1}b^n\right).
\ee
Since the $\mathbb{Z}_2$ automorphism and the killing form mutually are independent, we have,

\be
{\mathcal P}\left(f_0^{-n}\right)\equiv\Omega\left(f_0^{-n}\right)2\kappa\left[\Omega\left(g\right)\Omega\left(b^{-1}\right)\Omega\left(g^{-1}\right)\Omega\left(b^{-n}\right)\right]=f_0^{-n}.
\ee
In a similar way one can obtain ${\mathcal P}\left(f_{1,2}^{-n}\right)=\pm f_{1,2}^{-n}$ for all $n$s. Finally the curvature satisfies: $\Omega\left(\bar{F}_{tx}\right)=\bar{F}_{tx}$ which is a preserved property over the rotation given the anomaly function is parity-odd: ${\mathcal P}\left({\mathcal X}\right)=-{\mathcal X}$. The odd parity of the anomaly, apart from being consistent with the parity properties of the fields $(A,\,B)$ obtained already\footnote{See Eq. \ref{28}.}, immediately leads to asymptotic quasi-conservation of all the charges. To see this, integrating Eq. \ref{27} we obtain,

\be
Q^{-n}\left(\tilde{t},\,\tilde{x}\right)-Q^{-n}\left(-\tilde{t},\,-\tilde{x}\right)=\int_{-\tilde{t}}^{\tilde{t}}\int_{-\tilde{x}}^{\tilde{x}}{\mathcal X}f_0^{-n}\equiv 0,\label{QC1}
\ee
since the integrand is odd under space-time reflection. Herein the boundary points $\left(\pm\tilde{x},\,\pm\tilde{t}\right)$ can very well reach the infinity. Eq. \ref{QC1} is the very statement of quasi-conservation since all the charges are conserved asymptotically \cite{A3,A1}. In particular, going back to Eq.s \ref{N02} and \ref{NN03}, the charges $Q^{-1,-3,-4}$ are conserved at space-time boundaries subjected to the parity properties obtained for ${\mathcal X}$ and $A_\pm$ above. Therefore given the solutions $\left(A,\,B\right)$ with correct parity properties, {\it i. e.} ${\mathcal P}\left(A_{R,I}\right)=\pm A_{R,I}$ and ${\mathcal P}\left(B\right)=B$, quasi-conservation of the AB system can be obtained in a straight-forward manner. In particular cases, when the product ${\mathcal X}f_0^{-n}$ for some $n$ is odd under {\it space-reflection}, clearly the corresponding charge is conserved even locally. However it is not always easy to obtain solutions to such deformed systems \cite{tB,A1,OwnKdV}, though one can be hopeful since physical baroclinic systems are known to support localized solutions with definite parity \cite{Hart}. In the following we try to obtain a few of such solutions.

\subsection{Localized solutions and particular results}
Although QID need not be perturbative \cite{A3} one is allowed to consider an order-by-order expansion in the corresponding parameter $\varepsilon$ \cite{A3,A1}. The quasi-deformed function $B$ in Eq. \ref{23} can then be expanded in powers of $\varepsilon$ as,

\bea
&&\frac{1-B}{16}\nonumber\\
&&=\frac{1-\tilde{B}}{16}-\frac{\varepsilon}{16}\left[1-\tilde{B}+\left(\sqrt{2}-\sqrt{1+\tilde{B}}\right)^2\ln\left(\frac{\sqrt{2}-\sqrt{1+\tilde{B}}}{2\sqrt{2}}\right)\right]+{\mathcal O}\left(\varepsilon^2\right)\nonumber\\
&&=\frac{1-\tilde{B}}{16}-\frac{\varepsilon}{16}\left[1-B_0+\left(\sqrt{2}-\sqrt{1+B_0}\right)^2\ln\left(\frac{\sqrt{2}-\sqrt{1+B_0}}{2\sqrt{2}}\right)\right]\nonumber\\
&&\qquad+{\mathcal O}\left(\varepsilon^2\right).\label{39}
\eea
Inside the bracket $\tilde{B}$ was replaced by the undeformed counterpart $B_0$ maintaining a first-order expansion in $\varepsilon$. This is acceptable as $\tilde{B}=B_0+{\mathcal O}(\varepsilon)$ following which $\Psi=\psi+{\mathcal O}(\varepsilon)$. To obtain a $\varepsilon$-expansion of $B$, since $\tilde{B}=\cos\Psi$ and $A=\Psi_x$, it is safe to assume that the system $(A,\tilde{B})$ is closer to the undeformed system $(A_0,B_0)$ than $(A,B)$ is. Then, on considering the expansion $A=A_0+\varepsilon A_{(1)}+\cdots$, the first of Eq.s \ref{1} can approximately holds up, as the first order contribution to $\tilde{B}$ is considered to be complemented by the same in $A$. Thus,

\bea
&&\tilde{B}=-\frac{1}{2}\partial_x^{-1}\left(\vert A_0+\varepsilon A_{(1)}\vert^2\right)_t+{\mathcal O}\left(\varepsilon^2\right)\nonumber\\
&&\quad= B_0-\frac{\varepsilon}{2}\partial_x^{-1}\left(A_0A_{(1)}^*+A_0^*A_{(1)}\right)_t+{\mathcal O}\left(\varepsilon^2\right).\label{40}
\eea
Substituting these expressions in Eq. \ref{28} the ${\mathcal O}(\varepsilon)$ contribution to the QID anomaly has the form,

\bea
&&{\mathcal X}=\varepsilon{\mathcal X}^1+\varepsilon^2{\mathcal X}^2+\cdots\nonumber\\
&&\quad=-i\frac{\varepsilon}{4}\left[1-B_0+\left(\sqrt{2}-\sqrt{1+B_0}\right)^2\ln\left(\frac{\sqrt{2}-\sqrt{1+B_0}}{2\sqrt{2}}\right)\right]_x\nonumber\\
&&\qquad\quad+{\mathcal O}\left(\varepsilon^2\right),
\eea
which is free from the first-order deformation contribution $A_{(1)}$ and thus can be computed using undeformed solution $B_0$. Accordingly, Eq. \ref{40} is validated as the anomalous contributions can be obtained order-by-order in $\varepsilon$ in principle. In particular the ${\mathcal O}(\varepsilon)$ contribution is a total derivative\footnote{It is easy to see that up to all perturbative orders ${\mathcal X}$ is a total derivative given the first of Eq.s \ref{1} is valid at that order.} and thus $Q^{-1}$ above is trivially conserved subjected to proper boundary conditions. More importantly for the undeformed solution $B_0$ being even under parity, the first order contribution to the anomaly is odd.  

To determine the first order contributions to $A$ and $B$, from Eq.s \ref{39} and \ref{40} followed by $A_{xt}=AB$, we obtain,

\bea
&&B= B_0+\varepsilon\Bigg[1-B_0+\left(\sqrt{2}-\sqrt{1+B_0}\right)^2\ln\left(\frac{\sqrt{2}-\sqrt{1+B_0}}{2\sqrt{2}}\right)\nonumber\\
&&\qquad\qquad\qquad\quad-\frac{1}{2}\partial_x^{-1}\left(A_0A_{(1)}^*+A_0^*A_{(1)}\right)_t\bigg]+{\mathcal O}\left(\varepsilon^2\right),\nonumber\\
&&\left(\frac{\partial^2}{\partial x\partial t}-B_0\right)A_{(1)}= A_0\left[1+\left(\sqrt{2}-\sqrt{1+B_0}\right)^2\ln\left(\frac{\sqrt{2}-\sqrt{1+B_0}}{2\sqrt{2}}\right)\right].\label{N01}
\eea
So it boils down to solving the second equation above for $A_{(1)}$ which is a second order inhomogeneous equation in terms of the undeformed functions obtained by comparing ${\mathcal O}\left(\varepsilon\right)$ terms. Even with the simplest undeformed solutions the above equation may not be easy to solve exactly, though on principle the deformed solutions can thus be obtained completely one order at a time. On considering the undeformed single-soliton solutions of Eq.s \ref{NH01}, both parity-even with $A_0$ being real, which exactly meets the required conditions obtained above, the obvious sine-Gordon mapping validates the subsequent QID scheme. As $B_0=1-2{\rm sech}^2\theta$ is parity-even, the first-order contribution to the anomaly ${\mathcal X}^1$ is now explicitly parity-odd:

\bea
&&{\mathcal X}^1=-\frac{\gamma}{2}\left[3-\frac{1}{\sqrt{1-{\rm sech}^2\theta}}+\frac{2-2\sqrt{1-{\rm sech}^2\theta}}{\sqrt{1-{\rm sech}^2\theta}}\ln\left(\frac{1-\sqrt{1-{\rm sech}^2\theta}}{2}\right)\right]\nonumber\\
&&\qquad\quad\times{\rm sech}^2\theta\tanh\theta.
\eea
As a result the asymptotic conservation of the charges at ${\mathcal O}(\varepsilon)$ is ensured from Eq. \ref{27} since the ${\mathcal O}(0)$ contribution to $\{f_0^{-n}\}$ depends only on the undeformed solution $A_0=2i\gamma{\rm sech}\theta$ which is parity-even. However, these charges are not conserved locally as $\left\{dQ^{-n}/dt\right\}$ will have non-trivial time dependence in general, as directly can be verified from the Eq.s \ref{N02} and \ref{NN03}. So the system is seen to be quasi-integrable.

\begin{figure}
\centering
\begin{subfigure}[b]{0.45\textwidth}
\centering
\includegraphics[width=\textwidth]{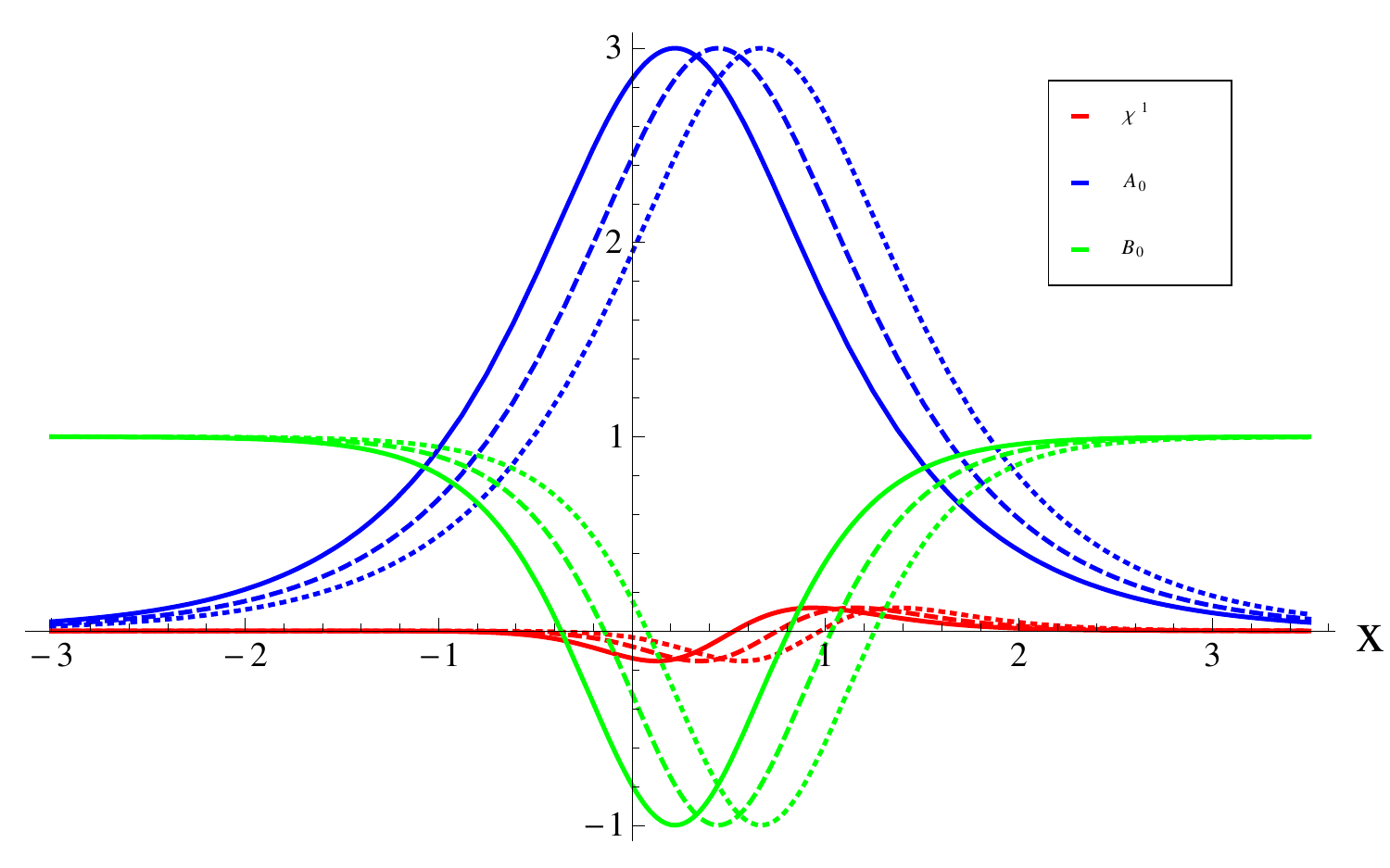}
\caption{\small{Undeformed 1-solitons and QID anomaly.}}
\label{F1}
\end{subfigure}
\begin{subfigure}[b]{0.45\textwidth}
\centering
\includegraphics[width=\textwidth]{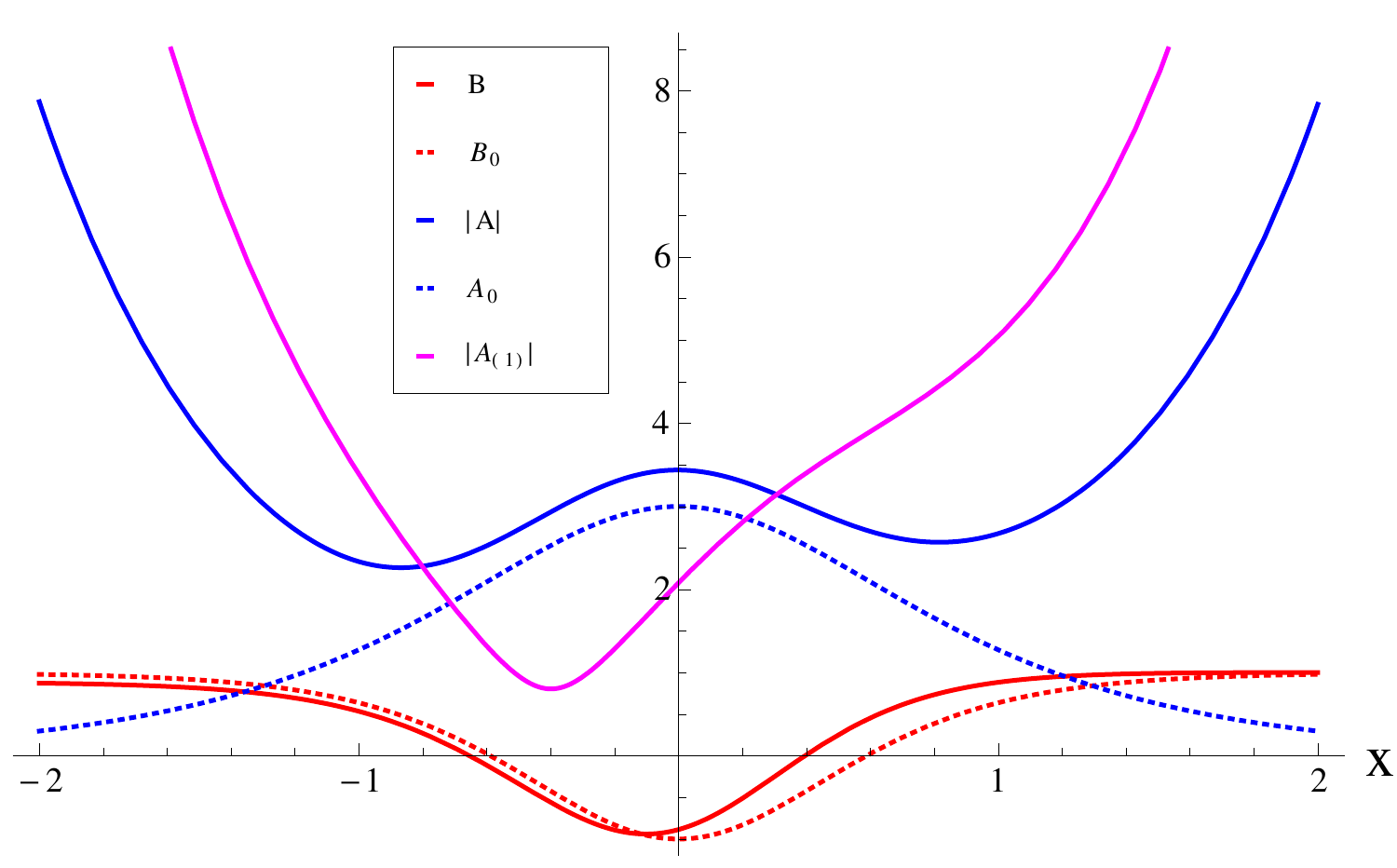}
\caption{Deformed 1-soliton amplitudes.}
\label{F2}
\end{subfigure}
\caption{a) The evolution of the parity-odd first-order anomaly ${\mathcal X}^{1}$ (red) along-with the undeformed parity-even single-soliton amplitudes $A_0$ (blue) and $B_0$ (green) as functions of position $x$ with $t=0.5$ (solid), $t=1$ (dashed) and $t=1.5$ (dotted). Here, $i\gamma=1.5$, $\delta=0$ and $\varepsilon=0.5$ are considered for optimum demonstration. The perturbative approach is justified as both $A_0$ and $B_0$ dominate ${\mathcal X}^1$. \\
b) The modulus of the quasi-perturbation $A_{(1)}$ (magenta) is found to be localized but asymmetric. Subsequently, the deformed amplitudes ($A$ (blue) and $B$ (red)) gets just marginally deformed from the undeformed counterparts ($B_0$ (red, dashed) and $A_0$ (blue, dashed)) at the temporal origin (t=0).}
\end{figure}

The behavior of the first-order anomaly contribution ${\mathcal X}^{1}$ against the undeformed solutions is depicted in Fig \ref{F1} which is clearly parity-odd. The anomaly is further subdominant against the undeformed amplitudes for $\varepsilon\sim 1$. Therefore the present approach of order-by-order expansion is justified in this case and contributions at higher orders can be evaluated in a successive manner.

The first order correction $A_{(1)}$, from Eq. \ref{N01}, to this single soliton case is defined by the following equation,

\bea
&&\left(\frac{\partial^2}{\partial x\partial t}-1+2{\rm sech}^2\theta\right)A_{(1)}\nonumber\\
&&\qquad\qquad= 2i\gamma{\rm sech}\theta\left[1+2\left(1-\vert\tanh\theta\vert\right)^2\ln\left(\frac{1-\vert\tanh\theta\vert}{2}\right)\right],\label{NN02}
\eea
with the homogeneous part being just $A_0$. We solve this equation graphically using Mathematica8, and depict the comparative results in Fig. \ref{F2}. The correction function $A_{(1)}$ is fairly localized but not even under space-reflection. As the amplitudes are only slightly deformed from the undeformed structure having correct space-time parity, the system may still maintain asymptotic integrability owing to the very local nature of such deformations. In a perturbative limit, following Eq. \ref{N01} the amplitudes $(A_0\to A,\,B_0\to B)$ deform only moderately, suggesting approximate conservation even locally.

We have seen that at ${\mathcal O}(\varepsilon)$ the charges do vanish asymptotically given the undeformed solution does and thus the system is quasi-integrable. As all other corrections must belong to higher-order contributions this result is exact. For the particular case of perturbative deformation, {\it i. e.}, when $\varepsilon\ll 1$ this analysis suggests absolute quasi-integrability. However at higher orders one needs to consider first-order deformations to the solutions which need not posses definite parities, as is evident from the graphical representation of $A_{(1)}$. As a result the term ${\mathcal X}f_0^n$ may not be parity odd any more at that order. Since the second of Eq.s \ref{N01} is itself parity-even one may obtain exact parity-even solution $A_{(1)}$ through more detailed numerical tools which is not within the scope of this paper. 

\begin{figure}
\centering
\begin{subfigure}[b]{0.45\textwidth}
\centering
\includegraphics[width=\textwidth]{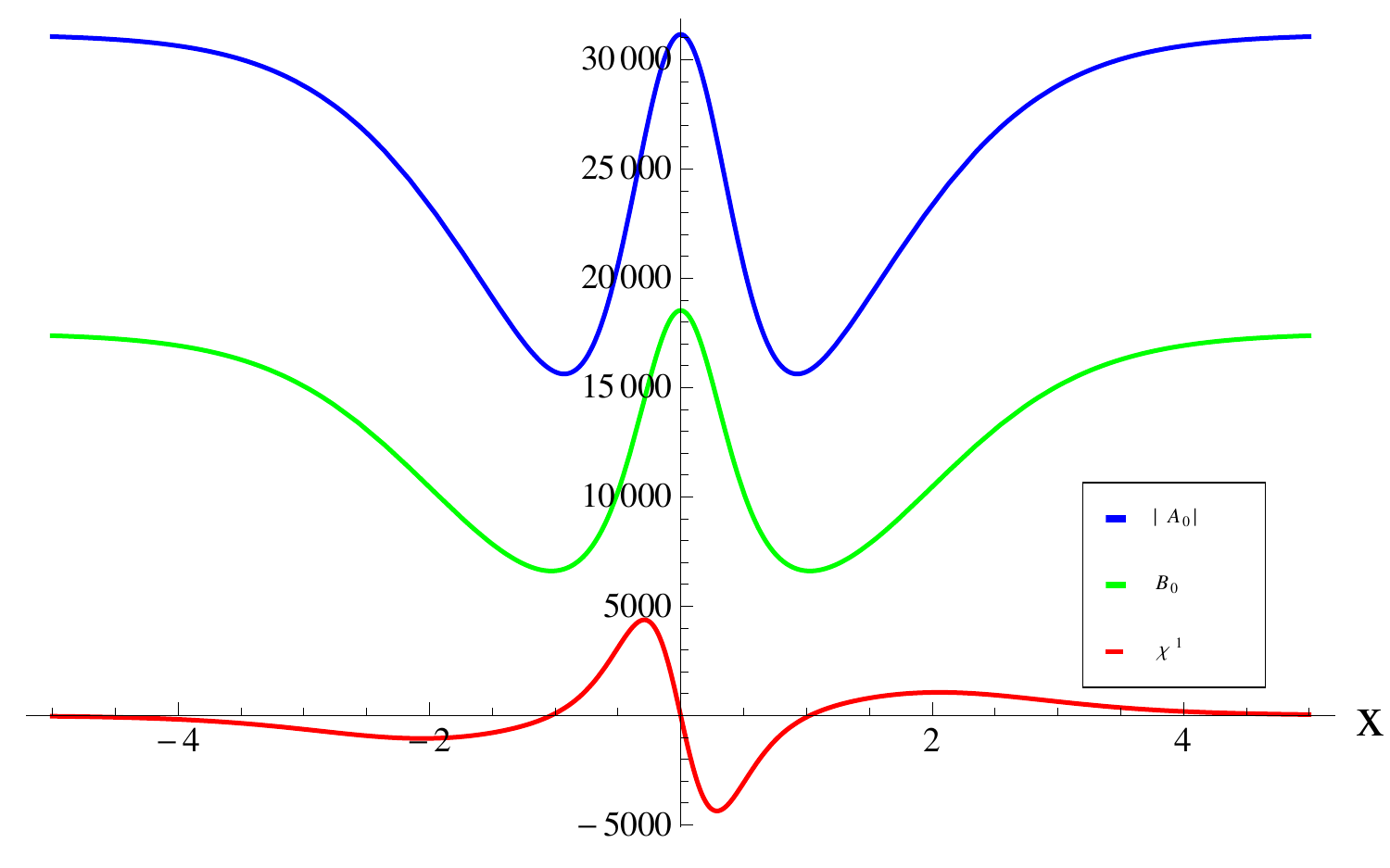}
\caption{Undeformed 2-solitons and QID anomaly.}
\label{F3}
\end{subfigure}
\begin{subfigure}[b]{0.45\textwidth}
\centering
\includegraphics[width=\textwidth]{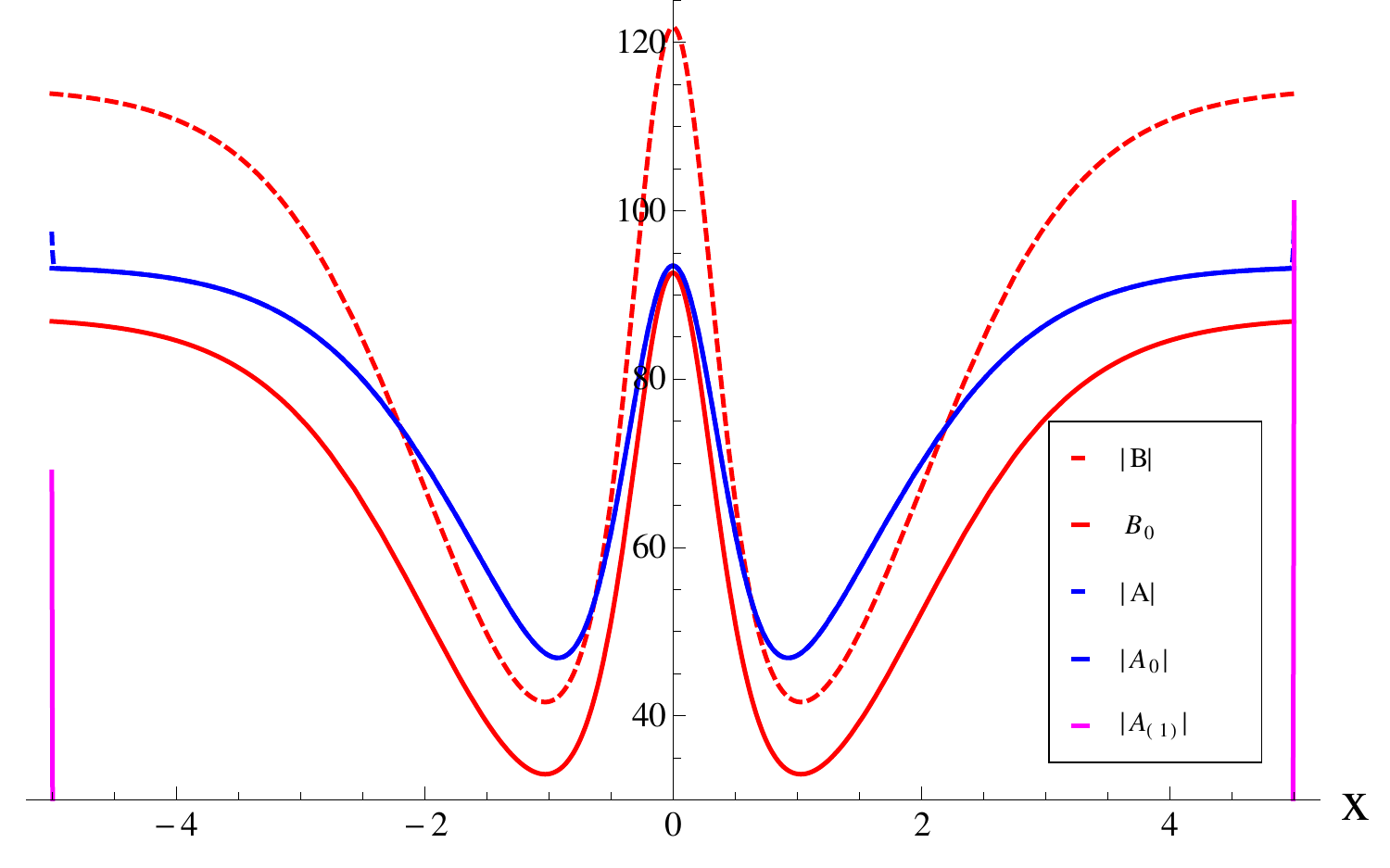}
\caption{Deformed 2-soliton amplitudes.}
\label{F4}
\end{subfigure}
\caption{a) The first-order anomaly ${\mathcal X}^1$ (red) for the two-soliton solution is both parity-odd and strictly localized, respecting the local symmetry of the undeformed amplitudes $\vert A_0\vert$ (blue) and $B_0$ (green) at the temporal origin. For a permissible perturbative range ($\varepsilon=0.1$) the anomaly is subdominant. Here, $a_1=1.1$, $a_2=1$ and $\delta_{1,2}=0$. \\
b) The first-order quasi-deformation to the modulus $\vert A_{(1)}\vert$ (magenta) is a steep well. The local amplitudes are effected minimally for a suitably small $\varepsilon$. The deformed amplitude $\vert A\vert$ (blue, dashed) subsequently differs from undeformed $\vert A_0\vert$ (blue) {\it only} at the boundary of the well. Deviation of the `potential' $\vert B\vert$ (red, dashed), though relatively prominent, maintains the local structure of $B_0$ (red). The amplitudes are scaled suitably for demonstration purpose.}
\end{figure}

We next consider QID of the two soliton case as depicted in Eq.s \ref{N03}. Here $A_0$ can either be real or imaginary whereas $B_0$ is still real and thus the `generalized' correspondence of Eq.s \ref{N1} is assumed for a sensible deformation. However, both the amplitudes are still parity-even as required. These amplitudes, along with the corresponding first order anomaly ${\mathcal X}^1$, have been depicted in Fig. \ref{F3} at temporal origin. The resultant anomaly is a parity-odd function as needed and is indeed subdominant for $\varepsilon=0.5$ to the 2-soliton system. The corresponding solutions to the Eq.s \ref{N01} are obtained numerically and the modified AB amplitudes are plotted in Fig \ref{F4}. The first order correction $A_{(1)}$ for the two-soliton case has a steep but localized well-like structure and is an even function. For $\varepsilon\to0$, $A_{(1)}$ respects the 2-soliton structure locally ($x\to0$). Consequently the deformed solitons $A$ and $B$ maintain similar local structures up to ${\mathcal O}(\varepsilon)$ with correct parity. Therefore the characteristic quasi-integrability should prevail in the 2-soliton case too. In principle, one can extend this procedure for higher-order multi-soliton solutions. However the actual computations will become successively tedious, even numerically.

The charges $Q^{-1,\,-2}$ are trivially conserved whereas the other charges are conserved asymptotically following the parity properties of the solutions and thus that of the anomaly.
Indeed, the system will lose quasi-integrability as one considers more complicated solution regimes wherein the parity properties are violated. Such cases include interacting solitonic structures and extended solutions since only localized and well-separated solitonic structures satisfy quasi-conservation \cite{A3,tB,A2,A1,A7}. The fact that we could obtain localized structures for the deformed system strongly suggests multiple conservation laws. A direct stability analysis of such solutions belonging to the deformed system can also explicitly show partial integrability given one can solve these systems directly. However, complete solutions are available only numerically for known quasi-integrable systems \cite{A3,tB,A2,A1,A7} which is beyond the present scope of study.

\section{Discussion and Conclusion}
We have analyzed and explicitly obtained both non-holonomic and quasi-integrable deformations of the AB system. In case of NHD the system develops nontrivial local inhomogeneities that satisfy higher order differential constraints. Such higher order constraints restrict the solution space of the deformed system yet do not effect the modified dynamics. The particular spectral order of the deformation to the temporal Lax operator is crucial for a meaningful result. A deformation of spectral order higher than $\lambda^{-1}$ will not effect the dynamics though increasingly more extended differential constraints can appear at lower spectral orders, which is a generic nature of the NHD itself. The amplitude $B_0$, when deformed, enjoys an arbitrariness in terms of a linear but local shift by the deformation parameter $u_1$ at order $\lambda^{-1}$ which can map between different undeformed solutions as well as can be utilized for achieving desired deformed solutions. 

We have obtained the deformed single-soliton solution akin to the undeformed counterpart, whereas a couple of known undeformed solutions of the kink-kink and kink-anti-kink types do not have deformed counterparts. Further a particular kink-type deformed solution was obtained not known for the undeformed AB system. Subsequent deforming local inhomogeneous functions $u_2,\,v_2$ are also obtained, which can model non-trivial local source terms occurring in realistic baroclinic fluids, and determines the particular deformed equations. These non-holonomic equations are `original integrable systems' with source, and are completely integrable. The more general approach of independently choosing the sources $u_2,\,v_2$ in principle allows for more accurate modelling of the baroclinic systems in real physical situations. Experimentally, baroclinic systems mostly deal with single-soliton structure of ${\rm sech}^2$-type in two-layer systems \cite{Huttemann,Hutter}. These `primary waves' or `1st modes' can very well be modelled by a NHD of the AB system in presence of different submarine structures \cite{Hutter,Chen2009}, the latter serving as local source terms. However, as is the case with NHD, the new integrable system may represent physical situations very distinct to that corresponding to the AB system, which warrants further studies.

On the other hand, the QID of the AB system corresponds to deformation of the potential (nonlinear) term that directly effects the nonlinearity. As a result only a subset of the total charges remain locally conserved, though all of them regain conservation asymptotically for solutions with correct parity. We have explicitly obtained initial few of those charges directly in terms of the dynamical variables, with a few of them being conserved locally. The $\mathbb{Z}_2$ symmetry analysis of the system \cite{A3,A2,A1,A4} entrusts definite parity properties with the variables making the anomalous charges to be asymptotically conserved in general, thereby quasi-integrability of the system is generally assured without subjecting to particular solutions. The few locally conserved charges correspond to certain basic symmetries of the system retained over QID; the identical conservation of $Q^{-2}$ may be related to the arbitrariness of $B_0\to B_d-u_1(x,t)$. 

The particular quasi-deformed single-soliton obtained at order $\varepsilon$ could model the experimentally observed single-soliton-like structures in bi-layer baroclinic waves \cite{Huttemann,Hutter} or the first-mode solitons modified in presence of bottom-lying obstacles in the fluid \cite{Chen2009}. Similar deformed soliton-structures have been obtained as internal solitary waves against a sloped bottom \cite{Qian2020,Zhao2004}. Breaking of such internal solitary waves against various obstacles \cite{Maderich} and their stratified mixing over submarine ridges \cite{Chen2007} have been observed, all supporting deformed single-soliton structures similar to what we have obtained. Though the experimental observation of two-soliton structures in two- or many-layer baroclinic system is challenging due to relatively smaller excitation energies \cite{Hutter}, the two-soliton quasi-deformation obtained herein could still represent physical situations as obstacles cause dissipation \cite{Chen2009,Maderich,Chen2007} which marks deviation from integrability. 

By keeping the order of the constraint equations fixed, the NHD can be extended to multiple integrable models over a number of higher spectral orders \cite{AK1,AK2} which are otherwise uncorrelated\footnote{For example, Ref. \cite{A8} and references therein.}. Such higher order NHDs of the AB system can be important in fluid dynamics and in non-linear optics. Possibility of QID for deformed solutions without the desired parity properties can also lead to further almost-integrable systems depicting real baroclinic and other fluid models. Both NHD and QID can be extended to multi-component AB systems \cite{Geng,Xie,Xu,Su,Zhang2021}, that generalizes the two-level baroclinic systems, for a better representation of the observed phenomena \cite{Huttemann,Hutter,Maderich}. Further, behavior of the AB rogue waves \cite{Wu,Wang2,Wang434,Geng,Su} under QID and NHD could be another fruitful study. Additionally, a study of the interaction dynamics of the deformed localized solutions can shed light on more complicated behavior of such systems. We hope to explore few of these aspects in the near future.

\section*{CRediT authorship contribution statement}
{\bf Kumar Abhinav}: Wrote the paper, performed all the calculations, made the plots, interpreted the results and did the physical analysis. {\bf Indranil Mukherjee}: Made the initial proposal, performed the initial calculations of the NHD part, gave physical inputs. {\bf Partha Guha}: Provided the initial motivation, supervised the writing and other aspects, provided various crucial insights and contributed to the introduction.  

\section*{Declaration of competing interest}
The authors declare that there is no competing financial and/or personal interests influencing the work reported in this manuscript.

\section*{Acknowledgement}
Kumar Abhinav's research is supported by Mahidol University under the grant numbered MRC-MGR 04/2565. Work of Partha Guha was supported by the Khalifa University of Science and
Technology under grant number FSU-2021-014.

\appendix
\section{Coefficients of rotated Lax components and curvature}

To evaluate the rotated spatial Lax operator $\bar{L}=gLg^{-1}++g_xg^{-1}$, on employing the 
Baker–Campbell–Hausdorff (BCH) formula, the expansions for the terms are obtained as:

\bea
&&gLg^{-1}=L+\sum_n\left[J_{-n},\,L\right]+\frac{1}{2!}\sum_{m,n}\left[J_{-m},\,\left[J_{-n},\,L\right]\right]\nonumber\\
&&\qquad\qquad\qquad+\frac{1}{3!}\sum_{l,m,n}\left[J_{-l},\,\left[J_{-m},\,\left[J_{-n},\,L\right]\right]\right]+\cdots;\nonumber\\
&&{\rm and}\nonumber\\
&&g_xg^{-1}=\sum_nJ_{-n,x}+\frac{1}{2!}\sum_{m,n}\left[J_{-m},J_{-n,x}\right]+\frac{1}{3!}\sum_{l,m,n}\left[J_{-l},\,\left[J_{-m},J_{-n,x}\right]\right]\nonumber\\
&&\qquad+\cdots
\eea
The first few of the terms are evaluated as follows:
\bea
&&\sum_n\left[J_{-n},\,L\right]\nonumber\\
&&=\sum_n\Big[\frac{\kappa}{2\sqrt{2}}\left(a_1^{-n}A_--a_2^{-n}A_+\right)b^{-n}+2i\left(a_1^{-n}F_2^{1-n}+a_2^{-n}F_1^{1-n}\right)\Big],\nonumber\\
&&\frac{1}{2!}\sum_{m,n}\left[J_{-m},\,\left[J_{-n},\,L\right]\right]\nonumber\\
&&=\sum_{m,n}\Big[i\kappa\left(a_1^{-m}a_1^{-n}-a_2^{-m}a_2^{-n}\right)b^{1-m-n}-\frac{\kappa}{2\sqrt{2}}\left(a_1^{-n}A_--a_2^{-n}A_+\right)\nonumber\\
&&\qquad\quad\times\left(a_1^{-m}F_2^{-m-n}+a_2^{-m}F_1^{-m-n}\right)\Big],\nonumber\\
&&\frac{1}{3!}\sum_{l,m,n}\left[J_{-l},\,\left[J_{-m},\,\left[J_{-n},\,L\right]\right]\right]\nonumber\\
&&=\sum_{l,m,n}\Big[-\frac{\kappa^2}{6\sqrt{2}}\left(a_1^{-n}A_--a_2^{-n}A_+\right)\left(a_1^{-l}a_1^{-m}-a_2^{-l}a_2^{-m}\right)b^{-l-m-n}\nonumber\\
&&\qquad\quad-i\frac{2\kappa}{3}\left(a_1^{-m}a_1^{-n}-a_2^{-m}a_2^{-n}\right)\left(a_2^{-l}F_1^{1-l-m-n}+a_1^{-l}F_2^{1-l-m-n}\right)\Big],\nonumber\\
&&\vdots\nonumber
\eea
wherein $A_\pm=\frac{A}{\kappa}\pm A^*$ and
\bea
&&\sum_nJ_{-n,x}=\sum_n\left(a_{1,x}^{-n}F_1^{-n}+a_{2,x}^{-n}F_2^{-n}\right),\nonumber\\
&&\frac{1}{2!}\sum_{m,n}\left[J_{-m},J_{-n,x}\right]=\sum_{m,n}\frac{\kappa}{2}\left(a_1^{-m}a_{2,x}^{-n}-a_2^{-m}a_{1,x}^{-n}\right)b^{-m-n},\nonumber\\
&&\frac{1}{3!}\sum_{l,m,n}\left[J_{-l},\left[J_{-m},J_{-n,x}\right]\right]\nonumber\\
&&=-\sum_{l,m,n}\frac{\kappa}{3}\left(a_1^{-m}a_{2,x}^{-n}-a_2^{-m}a_{1,x}^{-n}\right)\left(a_1^{-l}F_2^{-l-m-n}+a_2^{-l}F_1^{-l-m-n}\right),\nonumber\\
&&\frac{1}{4!}\sum_{k,l,m,n}\left[J_{-k},\,\left[J_{-l},\,\left[J_{-m},\,J_{-n,x}\right]\right]\right]\nonumber\\
&&=-\sum_{k,l,m,n}\frac{\kappa^2}{12}\left(a_1^{-m}a_{2,x}^{-n}-a_2^{-m}a_{1,x}^{-n}\right)\left(a_1^{-k}a_1^{-l}-a_2^{-k}a_2^{-l}\right)b^{-k-l-m-n},\nonumber\\
&&\vdots\label{15}
\eea
As $\bar{L}$ is independent of $F^n_{1,2}$, the resulting consistency conditions determine the expansion coefficients $a_{1,2}^{-n}$s. Starting from the highest order (coefficient of $F_{1,2}^{n=0}$) onward the first few of them are,

\bea 
&&a_{1,2}^{-1}=\frac{i}{4\sqrt{2}}A_\mp,\quad a_{1,2}^{-2}=-\frac{1}{8\sqrt{2}}A_{\pm,x},\nonumber\\
&&a_{1,2}^{-3}=-i\frac{1}{16\sqrt{2}}A_{\mp,xx}-i\frac{\kappa}{192\sqrt{2}}\left(A_+^2-A_-^2\right)A_\mp,\nonumber\\
&&a_{1,2}^{-4}=\pm\frac{\kappa}{96\sqrt{2}}A_\pm^2A_{\pm,x}\mp\frac{\kappa}{128\sqrt{2}}A_\mp^2A_{\pm,x}\mp\frac{\kappa}{384\sqrt{2}}A_{\mp,x}A_+A_-\nonumber\\
&&\qquad\quad+\frac{1}{32\sqrt{2}}A_{\pm,xxx},\nonumber\\
&&\vdots\label{16}
\eea
As a result the non-zero expansion coefficients of the rotated Lax component $\Bar{L}$, exclusively spanning the Kernel subspace, can be identified as,

\bea
&&\beta_L^1=-i,\quad\beta_L^0=0,\quad\beta_L^{-1}=-i\frac{\kappa}{32}\left(A_+^2-A_-^2\right),\nonumber\\
&&\beta_L^{-2}=-\frac{\kappa}{64}\left(A_{+,x}A_--A_{-,x}A_+\right),\nonumber\\
&&\beta_L^{-3}=i\frac{\kappa^2}{1536}\left(A_+^2-A_-^2\right)^2+i\frac{\kappa}{128}\left(A_+A_{+,xx}-A_-A_{-,xx}\right),\nonumber\\
&&\beta_L^{-4}=\frac{\kappa}{256}\left(A_{+,xxx}A_--A_{-,xxx}A_+\right)\nonumber\\
&&\qquad\quad+\frac{13\kappa}{12288}\left(A_+^2-A_-^2\right)\left(A_{+,x}A_--A_{-,x}A_+\right),\nonumber\\
&&\vdots,\label{17}
\eea

The rotated temporal Lax component $\Bar{M}=gMg^{-1}+g_tg^{-1}$ is evaluated next. The BCH expansions of the two terms in it are:

\bea
&&gMg^{-1}=M+\sum_n\left[J_{-n},\,M\right]+\frac{1}{2!}\sum_{m,n}\left[J_{-m},\,\left[J_{-n},\,M\right]\right]\nonumber\\
&&\qquad\qquad\qquad+\frac{1}{3!}\sum_{l,m,n}\left[J_{-l},\,\left[J_{-m},\,\left[J_{-n},\,M\right]\right]\right]+\cdots,\nonumber\\
&&{\rm and}\nonumber\\
&&g_tg^{-1}=\sum_nJ_{-n,t}+\frac{1}{2!}\sum_{m,n}\left[J_{-m},J_{-n,t}\right]+\frac{1}{3!}\sum_{l,m,n}\left[J_{-l},\,\left[J_{-m},J_{-n,t}\right]\right]+\cdots
\eea
Expressions for the first few of the commutators are found to be,

\bea
&&\sum_n\left[J_{-n},\,M\right]\nonumber\\
&&=-\sum_n\Big[i\frac{\kappa}{4\sqrt{2}}\left(a_1^{-n}A_{+,t}-a_2^{-n}A_{-,t}\right)b^{-1-n}+i\frac{B}{2}\left(a_1^{-n}F_2^{-1-n}+a_2^{-n}F_1^{-1-n}\right)\Big],\nonumber\\
&&\frac{1}{2!}\sum_{m,n}\left[J_{-m},\,\left[J_{-n},\,M\right]\right]\nonumber\\
&&=\sum_{m,n}\Big[-i\frac{\kappa B}{4}\left(a_1^{-m}a_1^{-n}-a_2^{-m}a_2^{-n}\right)b^{-1-m-n}\nonumber\\
&&\qquad\quad+i\frac{\kappa}{4\sqrt{2}}\left(a_1^{-n}A_{+,t}-a_2^{-n}A_{-,t}\right)\left(a_2^{-m}F_1^{-1-m-n}+a_1^{-m}F_2^{-1-m-n}\right)\Big],\nonumber\\
&&\frac{1}{3!}\sum_{l,m,n}\left[J_{-l},\,\left[J_{-m},\,\left[J_{-n},\,M\right]\right]\right]\nonumber\\
&&=\sum_{l,m,n}\Big[i\frac{\kappa^2}{12\sqrt{2}}\left(a_1^{-n}A_{+,t}-a_2^{-n}A_{-,t}\right)\left(a_1^{-m}a_1^{-l}-a_2^{-m}a_2^{-l}\right)b^{-1-l-m-n}\nonumber\\
&&\qquad\quad+i\frac{\kappa B}{6}\left(a_1^{-m}a_1^{-n}-a_2^{-m}a_2^{-n}\right)\left(a_1^{-l}F_2^{-1-l-m-n}+a_2^{-l}F_1^{-1-l-m-n}\right)\Big],\nonumber\\
&&\vdots\nonumber
\eea
and
\bea
&&\sum_nJ_{-n,t}=\sum_n\left(a_{1,t}^{-n}F_1^{-n}+a_{2,t}^{-n}F_2^{-n}\right),\nonumber\\
&&\frac{1}{2!}\sum_{m,n}\left[J_{-m},J_{-n,t}\right]=\sum_{m,n}\frac{\kappa}{2}\left(a_1^{-m}a_{2,t}^{-n}-a_2^{-m}a_{1,t}^{-n}\right)b^{-m-n},\nonumber\\
&&\frac{1}{3!}\sum_{l,m,n}\left[J_{-l},\left[J_{-m},J_{-n,t}\right]\right]\nonumber\\
&&=-\sum_{l,m,n}\frac{\kappa}{3}\left(a_1^{-m}a_{2,t}^{-n}-a_2^{-m}a_{1,t}^{-n}\right)\left(a_1^{-l}F_2^{-l-m-n}+a_2^{-l}F_1^{-l-m-n}\right),\nonumber\\
&&\frac{1}{4!}\sum_{k,l,m,n}\left[J_{-k},\,\left[J_{-l},\,\left[J_{-m},\,J_{-n,t}\right]\right]\right]\nonumber\\
&&=-\sum_{k,l,m,n}\frac{\kappa^2}{12}\left(a_1^{-m}a_{2,t}^{-n}-a_2^{-m}a_{1,t}^{-n}\right)\left(a_1^{-k}a_1^{-l}-a_2^{-k}a_2^{-l}\right)b^{-k-l-m-n},\nonumber\\
&&\vdots.\label{20}
\eea
Although the exact expressions of the expansion coefficients of $\Bar{M}$ are not crucial to determine the charges, we list a few of them for the sake of completion:

\bea
&&\beta_M^{-1}=i\frac{B}{4},\quad \beta_M^{-2}=\frac{\kappa}{64}\left(A_{+,t}A_--A_{-,t}A_+\right),\nonumber\\
&&\beta_M^{-3}=i\frac{\kappa}{128}\left(A_{+,x}A_+-A_{-,x}A_-\right)_t-i\frac{\kappa B}{128}\left(A_+^2-A_-^2\right),\nonumber\\
&&\beta_M^{-4}=\frac{\kappa}{256}\left(A_{+,xx}A_--A_{-,xx}A_+\right)_t+\frac{\kappa}{256}\left(A_{+,x}A_{-,xt}-A_{-,x}A_{+,xt}\right)\nonumber\\
&&\qquad\quad-\frac{\kappa B}{128}\left(A_{+,x}A_--A_{-,x}A_+\right)+\frac{\kappa^2}{12288}\left(A_+^2-A_-^2\right)\left(A_{+,t}A_--A_{-,t}A_+\right),\nonumber\\
&&\vdots\nonumber
\eea
\bea
&&\alpha_{1,2}^{-1}=0,\quad\alpha_{1,2}^{-2}=-\frac{1}{8\sqrt{2}}\left(A_{\pm,xt}-BA_\pm\right),\nonumber\\
&&\alpha_{1,2}^{-3}=-i\frac{1}{16\sqrt{2}}A_{\mp,xxt}+i\frac{B}{16\sqrt{2}}A_{\mp,x}-i\frac{\kappa}{192\sqrt{2}}\left(A_{+,t}A_--A_{-,t}A_+\right)A_\pm\nonumber\\
&&\qquad\quad-i\frac{\kappa}{192\sqrt{2}}\left\{\left(A_+^2-A_-^2\right)A_\mp\right\}_t,\nonumber\\
&&\alpha_{1,2}^{-4}=-\frac{B}{32}A_{\pm,xx}-\frac{\kappa B}{256\sqrt{2}}\left(A_+^2-A_-^2\right)A_\pm+\frac{\kappa}{256\sqrt{2}}\Big[8A_{\pm,xxxt}\pm 3A_{\pm,xt}A_\pm^2\nonumber\\
&&\qquad\quad\mp 2A_{\pm,xt}A_\mp^2\pm 6A_{\pm,x}A_{\pm,t}A_\pm\mp 2A_{\mp,x}A_{\mp,t}A_\pm\mp A_{\mp,xt}A_+A_-\nonumber\\
&&\qquad\quad\mp 4A_{\pm,x}A_{\mp,t}A_\mp\Big],\nonumber\\
&&\vdots\label{21a}
\eea

Finally the rotated curvature can be expressed as the following BCH series:

\bea
&&\Bar{F}_{tx}={\mathcal X}\sum_n\left(f_0^{-n}b^{-n}+f_1^{-n}F_1^{-n}+f_2^{-n}F_2^{-n}\right)\nonumber\\
&&\qquad={\mathcal X}\Bigg[b^{-1}+\sum_n\left[J_{-n},\,b^{-1}\right]+\frac{1}{2!}\sum_{m,n}\left[J_{-m},\,\left[J_{-n},\,b^{-1}\right]\right]\nonumber\\
&&\qquad\qquad\quad+\frac{1}{3!}\sum_{l,m,n}\left[J_{-l},\,\left[J_{-m},\,\left[J_{-n},\,b^{-1}\right]\right]\right]+\cdots\Bigg].\label{22a}
\eea
The first few of the individual commutators can readily be evaluated as,

\bea
&&\sum_n\left[J_{-n},\,b^{-1}\right]=-2\sum_n\left(a_1^{-n}F_2^{-1-n}+a_2^{-n}F_1^{-1-n}\right),\nonumber\\
&&\frac{1}{2!}\sum_{m,n}\left[J_{-m},\,\left[J_{-n},\,b^{-1}\right]\right]=-\sum_{m,n}\kappa\left(a_1^{-m}a_1^{-n}-a_2^{-m}a_2^{-n}\right)b^{-1-m-n},\nonumber\\
&&\frac{1}{3!}\sum_{l,m,n}\left[J_{-l},\,\left[J_{-m},\,\left[J_{-n},\,b^{-1}\right]\right]\right]\nonumber\\
&&=\sum_{l,m,n}\frac{2\kappa}{3}\left(a_1^{-m}a_1^{-n}-a_2^{-m}a_2^{-n}\right)\left(a_1^{-l}F_2^{-1-l-m-n}+a_2^{-l}F_1^{-1-l-m-n}\right),\nonumber\\
&&\vdots.\label{23a}
\eea
The expansion coefficients of the rotated curvature can be obtained as usual order-by-order as:

\bea
&&f_0^{-1}=1,\quad f_0^{-2}=0,\quad f_0^{-3}=-\frac{\kappa}{32}\left(A_+^2-A_-^2\right),\nonumber\\
&&f_0^{-4}=i\frac{\kappa}{32}\left(A_{+,x}A_--A_{-,x}A_+\right),\nonumber\\
&&\vdots\nonumber\\
&&f_{1,2}^{-1}=0,\quad f_{1,2}^{-2}=-i\frac{1}{2\sqrt{2}}A_\pm,\quad f_{1,2}^{-3}=\frac{1}{4\sqrt{2}}A_{\mp,x},\nonumber\\
&&f_{1,2}^{-4}=i\frac{1}{8\sqrt{2}}A_{\pm,xx}+i\frac{\kappa}{64\sqrt{2}}\left(A_+^2-A_-^2\right)A_\pm,\nonumber\\
&&\vdots\label{24a}
\eea


\medskip

\end{document}